\documentclass[aps,pre,superscriptaddress,twocolumn,longbibliography,groupedaddress]{revtex4-2}

\usepackage[colorlinks]{hyperref}
\hypersetup{
  linkcolor = {blue},
  citecolor = {blue},
  urlcolor = {cyan}
}

\usepackage{float}
\usepackage{graphicx}
\usepackage{caption}
\usepackage{subcaption}
\usepackage{amsmath}
\usepackage{amsthm,amssymb}
\usepackage{amsfonts}
\usepackage{booktabs}
\usepackage{siunitx}
\usepackage{halloweenmath}

\usepackage{bm}
\usepackage{epstopdf}
\usepackage{algorithm}
\usepackage{algpseudocode}
\usepackage{mathrsfs}
\usepackage{mathptmx}%
\usepackage[all]{xy}
\usepackage{pbox}
\usepackage{verbatim}
\usepackage{mathtools}
\usepackage{tikz,pgfplots}
\usepackage{xfrac}
\usepackage{cleveref}
\usepackage{url}
\usepackage{pict2e}
\usepackage{tikz,tikz-3dplot}
\usetikzlibrary{perspective,shapes}
\usepackage{xkeyval}
\usepackage{xcolor}
\usepackage[T1]{fontenc}
\usepackage{curve2e}
\usepackage{xtab,afterpage}
\usepackage{balance}
\usepackage[version=4]{mhchem}
\usepackage{braket}
\usepackage{physics}
\usepackage{tikz}
\usetikzlibrary{calc}

\newcommand{\er}[1]{Eq.~\eqref{#1}}

\newcommand{\era}[2]{Eqs.~(\ref{#1}) and (\ref{#2})}

\newcommand{\opcat}[1]{{#1}_{\mathrm{TPM}}}
\newcommand{\opcatq}[1]{{#1}_{\mathrm{qTPM}}}
\newcommand{\opcatz}[1]{{#1}_{\mathrm{TPMz}}}
\newcommand{\opcatqz}[1]{{#1}_{\mathrm{qTPMz}}}

\newcommand{\opcatpxp}[1]{{#1}_{\mathrm{PXP}}}
\newcommand{\opcatpppxppp}[1]{{#1}_{\mathrm{P_6X}}}

\newcommand{\opcatlgt}[1]{{#1}_{\mathrm{LGT}}}

\makeatletter

\@ifdefinable\SuCmathpictvertex{} 
\@ifdefinable\@SuC@reserved@dimen{\newdimen\@SuC@reserved@dimen}

\newenvironment*{@SuC@math@picture}[8]{%
  \def\SuCmathpictvertex{\circle*{#6}}%
  \setlength\unitlength{\fontdimen 22 #5\tw@}%
  \setlength\@SuC@reserved@dimen{#7\unitlength}%
  \kern\@SuC@reserved@dimen
  \@HwM@d@pict@strut{#2}%
  \picture(#3,#1)(#4,-1)%
    \roundcap
    \roundjoin
    \linethickness{#8\@HwM@thickness@units@for #5}%
}{%
  \endpicture
  \kern\@SuC@reserved@dimen
}
\newcommand*\@SuC@general@pict[9]{%
  \begin{@SuC@math@picture}%
            {#2}{#3}
            {#4}{#5}
            #6
            {#7}
            {#8}
            {#9}
    #1%
  \end{@SuC@math@picture}%
}
\newcommand*\@SuC@math@version@shunt[7]{%
  \@HwM@choose@thicknesses{\@SuC@general@pict {#1}{#2}{#3}{#4}{#5}#7}%
      {{.8}{.4}{}}
      {{1}{.5}{1.5}}
}

\newcommand*\DeclareNewSuCMathPict[6]{%
  \newcommand*{#1}{%
    \@HwM@general@ordinary@symbol
      {\@SuC@math@version@shunt {#6}{#2}{#3}{#4}{#5}}%
  }%
}

\makeatother

\DeclareNewSuCMathPict{\trigonleft}
            {3}{1}  
            {4}{-2} 
{
    \polygon(45:2)(135:2)(225:2)%
    \put (45:2){\SuCmathpictvertex}%
    \put(135:2){\SuCmathpictvertex}%
    \put(225:2){\SuCmathpictvertex}%
}

\DeclareNewSuCMathPict{\tetrapleuro}
            {3}{1}  
            {4}{-2} 
{
    \polygon(45:2)(135:2)(270:2)%
    \put (45:2){\SuCmathpictvertex}%
    \put (90:1.4){\SuCmathpictvertex}%
    \put(135:2){\SuCmathpictvertex}%
    \put(270:2){\SuCmathpictvertex}%
}

\DeclareNewSuCMathPict{\tetragon}
            {3}{1}  
            {4}{-2} 
{
    \polygon(45:2)(135:2)(225:2)(315:2)%
    \put (45:2){\SuCmathpictvertex}%
    \put(135:2){\SuCmathpictvertex}%
    \put(225:2){\SuCmathpictvertex}%
    \put(315:2){\SuCmathpictvertex}%
}

\DeclareNewSuCMathPict{\kagome}
            {3}{1}  
            {4}{-2} 
{
    \polygon(45:2)(135:2)(284:6)(256:5.5)(45:2)%
    \put (45:2){\SuCmathpictvertex}%
    \put (256:5.5){\SuCmathpictvertex}%
    \put(135:2){\SuCmathpictvertex}%
    \put (284:5.5){\SuCmathpictvertex}%
    \put(270:2){\SuCmathpictvertex}%
}

\begin{document}

\title{The quantum Newman-Moore model in a longitudinal field}

\author{Konstantinos Sfairopoulos}
\email{ksfairopoulos@gmail.com}
\affiliation{School of Physics and Astronomy, University of Nottingham, Nottingham, NG7 2RD, UK}
\affiliation{Centre for the Mathematics and Theoretical Physics of Quantum Non-Equilibrium Systems,
University of Nottingham, Nottingham, NG7 2RD, UK}
\author{Juan P. Garrahan}
\affiliation{School of Physics and Astronomy, University of Nottingham, Nottingham, NG7 2RD, UK}
\affiliation{Centre for the Mathematics and Theoretical Physics of Quantum Non-Equilibrium Systems,
University of Nottingham, Nottingham, NG7 2RD, UK}

\begin{abstract}
We study the quantum Newman-Moore model, or quantum triangular plaquette model (qTPM), in the presence of a longitudinal field (qTPMz). We present evidence that indicates that the ground state phase diagram of the qTPMz includes various frustrated phases breaking translational symmetries, dependent on the specific sequence of system sizes used to take the large-size limit. This phase diagram includes the known first-order phase transition of the qTPM, but also additional first-order transitions due to the frustrated phases. Using the average longitudinal magnetization as an order parameter, we analyze the magnetization plateaus that characterize the ground state phases, describe their degeneracies, and obtain the qTPMz phase diagram using classical transfer matrix and quantum matrix product state techniques. We identify a region of the parameter space which can be effectively described by a Rydberg blockade model on the triangular lattice. At the same time, it can be reformulated as an effective lattice gauge theory but also a model of quantum trimers. In this same region we provide a variational ground state wavefunction similar to a resonating valence bond solid state which accurately describes the ground state wavefunction of the system. Lastly, we fail to find indications of a phase with $\mathbb{Z}_2$ topological order connecting the quantum paramagnetic and classical frustrated phases. 
\end{abstract}

\maketitle

\section{Introduction}

\begin{figure}[t]
    \centering
    \includegraphics[width=0.6\linewidth]{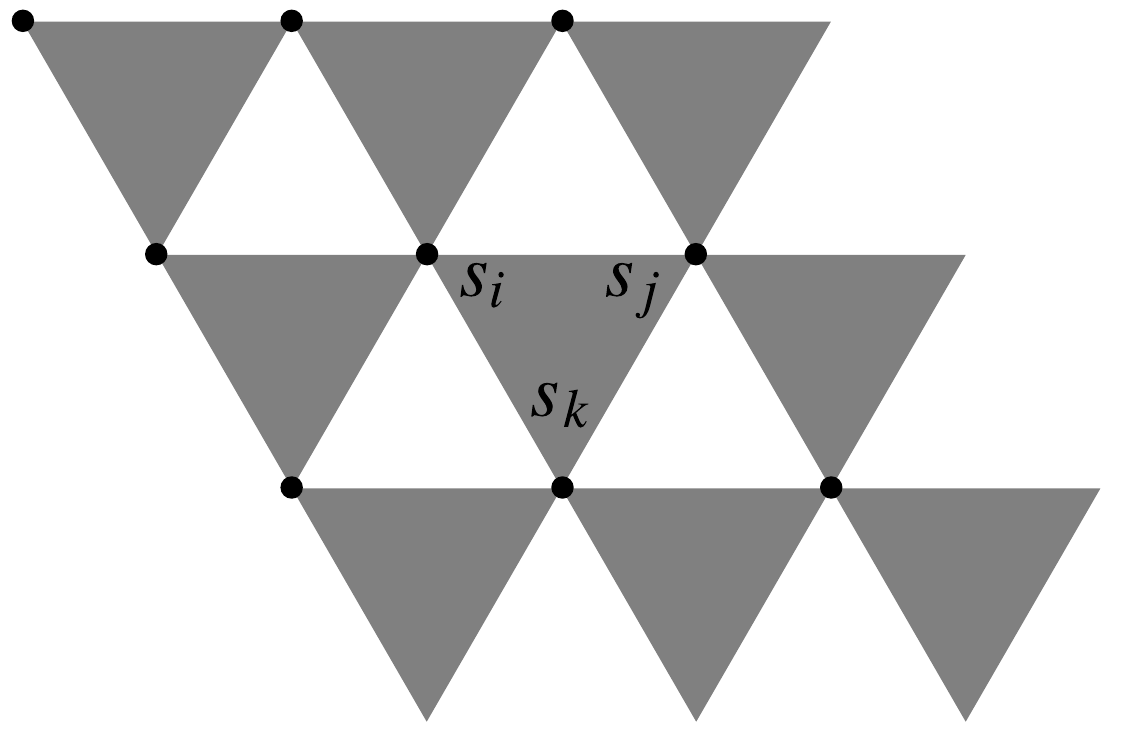}
    \caption{Geometry of the TPM. Spins reside in the sites of a triangular lattice. Interactions are between spins in the downward pointing plaquettes of the lattice. We show a system of size $N=3\times3$ with periodic boundary conditions.}
     \label{fig:TPM_lattice}
\end{figure}

The classical Newman-Moore model \cite{newman1999glassy,garrahan2000glassiness,garrahan2002glassiness}, also known as the triangular plaquette model (TPM), 
is a well-known classical spin system, originally studied as an example of an effective kinetically constrained model (KCM) \cite{ritort2003glassy}
in the context of the glass transition and dynamical facilitation theory \cite{chandler2010dynamics,speck2019dynamic}. 
The TPM
consists of Ising spins interacting in triplets on the downward-pointing triangles of the triangular lattice, with energy function
\cite{newman1999glassy,garrahan2000glassiness,garrahan2002glassiness} 
\begin{equation}{\label{TPMclassical}}
    \opcat{E} =  - J \sum_{\{i, j, k\} \in \triangledown} s_i s_j s_k ,
\end{equation} 
where $\{i, j, k\}$ indicate the sites of downward-pointing plaquettes, and $s_i = \pm 1 $; see Fig.~\ref{fig:TPM_lattice}.
From the point of view of glass physics \cite{biroli2013perspective:}, the TPM is interesting because it combines simple static behaviour in equilibrium with slow and heterogeneous relaxation~\cite{newman1999glassy,garrahan2000glassiness,garrahan2002glassiness}, without the need to impose explicit dynamical constraints as in KCMs \cite{ritort2003glassy}.

At zero temperature, the TPM with periodic boundary conditions (PBC) and at least one dimension a power of two has a unique ground state with all spins aligned \cite{newman1999glassy}. Furthermore, under these conditions an exact mapping to free defects exists, such that 
\begin{equation}
    \opcat{E} = - J \sum_{\triangledown} d_\triangledown , 
    \label{plaq}
\end{equation}
with the mapping between defects and spins, $d_\triangledown = s_i s_j s_k$, being bijective. Since the energy, Eq.~\eqref{plaq}, is that of a free gas of binary defects, it follows that the static properties of the TPM with PBC are trivial (i.e, disordered and with no defect correlations) at all temperatures. In contrast, the dynamics of the defects is constrained \cite{garrahan2000glassiness,garrahan2002glassiness}, which gives rise to slow glassy relaxation at low temperatures despite the simplicity of its thermodynamic properties. 

For other boundary conditions, the ground state structure of \er{TPMclassical} can be more intricate \cite{sfairopoulos2023boundary}. This stems from a connection between the minima of \er{TPMclassical} and the limit cycles of a one-dimensional cellular automaton (CA), specifically, for the TPM, Wolfram's rule 60 \cite{wolfram1983statistical,martin1984algebraic}. Using this connection Ref.~\cite{sfairopoulos2023boundary} systematically obtained the classical ground states of the TPM for different boundary conditions and system size sequences. 

Adding a transverse field to \er{TPMclassical} defines the Hamiltonian of the quantum triangular plaquette model (qTPM) \cite{yoshida2014quantum,devakul2019classifying,vasiloiu2020trajectory,zhou2021fractal,inack2022neural,2024_Wiedmann}
\begin{equation}{\label{QTPM}}
    \opcatq{H} =  - J \sum_{\{i, j, k\} \in \triangledown} Z_i Z_j Z_k - h \sum_i X_i,
\end{equation}
with $Z_i$ and $X_i$ the Pauli operators acting nontrivially on site $i$, and with the classical configurations of \er{TPMclassical} corresponding to the basis states where $Z_i$ is diagonal. 
The qTPM has a ground-state transition that coincides with its self-dual point \cite{vasiloiu2020trajectory}, as predicted by standard duality arguments \cite{kramers1941statistics,cobanera2011the-bond-algebraic}.  Numerical evidence indicates that this transition is of first order for PBC \cite{yoshida2014quantum,devakul2019classifying,vasiloiu2020trajectory,sfairopoulos2023boundary}, including additional spontaneous symmetry breaking for sequences of system sizes which have degenerate classical ground states \cite{sfairopoulos2023boundary}. 

In this work we continue the exploration of the ground state phases of the qTPM by adding to the Hamiltonian of the qTPM a longitudinal field. We use for the resulting model the acronym qTPMz, 
\begin{equation}{\label{QTPMz}}
    \opcatqz{H} =  - J \sum_{\{i, j, k\} \in \triangledown} Z_i Z_j Z_k - h \sum_i X_i - k \sum_i Z_i , 
\end{equation}
and study systems with PBC and dimensions $N = L\times M$. Our central result is the ground state phase diagram of the qTPMz as a function of transverse and longitudinal field strengths, showing the existence of various frustrated phases which break translation symmetries and whose existence is sensitive to boundary conditions and to the manner in which the thermodynamic limit is approached. 

Interestingly, we observe that in a region of this phase diagram the system can be effectively described by a model displaying a Rydberg-like blockade on the triangular lattice. Recent years have seen the emergence of the Rydberg atoms platform as one of the primary candidates for studying low-energy or nonequilibrium quantum dynamics \cite{henriet2020quantum,browaeys2020many-body,wu2021a-concise,cheng2024emergent}. A much used model to decribe Rydberg atoms in optical traps is in terms of a spin-$1/2$ Ising model, see e.g.~\cite{weimer2008quantum,sun2008numerical,olmos2010thermalization} with interactions between excited atoms being of the van der Waals type (induced-dipole) for Rydberg $s$ states, $V_{ij}= C | r_i - r_j|^{-6}$. The rapid growth of this interaction with decreasing distance between excited atoms gives rise to the ``blockade'' phenomenon, whereby excitations within a distance, $R_{\rm b}$, from other excitations are effectively suppressed. A useful approximation is to replace the interaction by one with a sharp cut-off at $R_{\rm b}$, so that $V_{ij}=\infty$, if $|i-j|<R_{\rm b}$, and $V_{ij}=0$ otherwise. In one dimension and for $R_{\rm b}$ corresponding to the lattice spacing, the interaction Hamiltonian is given by the so-called PXP model
\cite{fendley2004competing,lesanovsky2011many-body,lesanovsky2012interacting,turner2018weak}
\begin{equation}{\label{PXP}}
   \opcatpxp{H} = \Omega \sum_i P_{i-1} X_i P_{i+1} - \Delta \sum_i Z_i . 
\end{equation}
where the operator $P_i = 1 - n_i$ projects out the excited state in site $i$. A pair of neighbouring excitations represents a violation of the blockade condition. Under \er{PXP} such pairs are immobile, and dynamics takes place in subspaces where their number remains constant and their location fixed. The largest of these subspaces is that with no excitation pairs. 

By employing a relation between the Rydberg description of our model and an emergent $\text{U}(1)$ lattice gauge theory (LGT) \cite{cheng2022tunable,cheng2023gauge,cheng2023variational,cheng2024emergent}, we also construct an LGT description of our system. At the same time, we provide a complementary one in terms of quantum trimers too. Then, we devise a variational ansatz, inspired by mean-field quantum spin liquid (QLS) models \cite{book_Wen} which proves to be accurate for describing the ground state of the model across its first-order quantum phase transitions between the paramagnetic and frustrated phases. Lastly, we fail to find indications of a spin liquid phase separating the quantum paramagnet with the crystal phases.

The paper is organized as follows. In order to get intuition about the model, we first study in Sec.~\ref{classical_frustration} the minima of the {\em classical} TPM in a field [or TPMz, obtained from \er{QTPMz} by setting $h=0$]. While this model was studied first in Ref.~\cite{sasa2010thermodynamic} (see also Refs.~\cite{garrahan2014transition,turner2015overlap}), we show that the competition between interaction and longitudinal field terms in its energy function gives rise to frustration and to an intricate plateau structure of their magnetizations. We build on these results in Sec.~\ref{QPT} to construct the expected ground state phase diagram of the qTPMz in terms of both transverse and longitudinal fields, as obtained from extensive numerical simulations. In Sec.~\ref{k=-3J} we consider the qTPMz in the parameter region of small transverse fields and at negative enough longitudinal field, between the frustrated phases and the quantum paramagnetic phase. We show that near this point the quantum system can be thought of as a spin model in a constrained state space subject to a ``Rydberg blockade'', with an equivalent description in terms of an effective lattice gauge theory or in terms of quantum {\em trimers}. In Sec.~\ref{sec:topo} we consider the further possibility that the quantum paramagnetic and frustrated phases are separated by a spin liquid phase. We conclude with a discussion in Sec.~\ref{conclusions}. Extra details are provided in the Appendices. 

\begin{figure}[t]
   \centering
   \begin{subfigure}[b]{\linewidth}
       \includegraphics[width=0.8\linewidth]{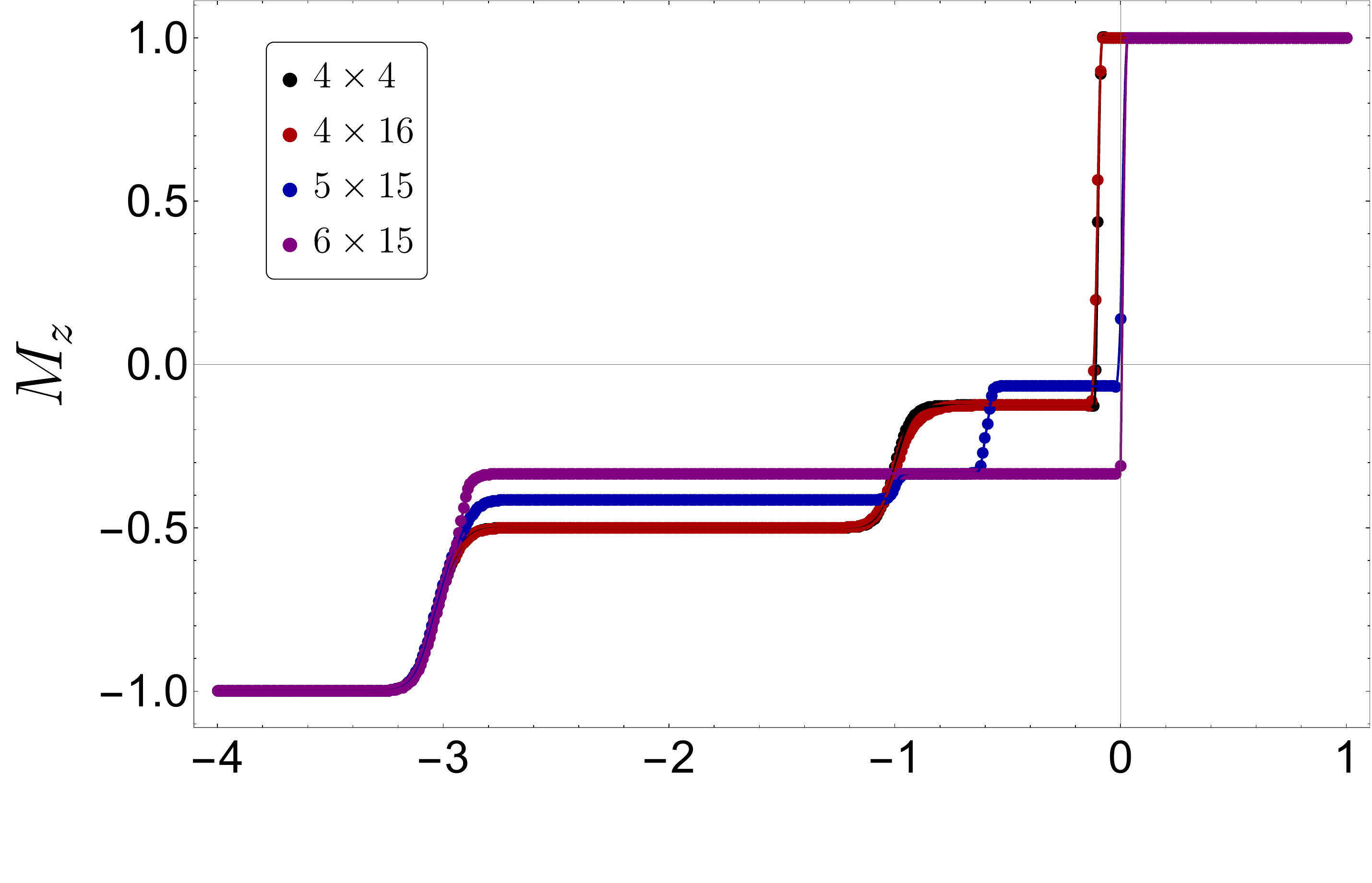}
    \end{subfigure}
    \begin{subfigure}[b]{\linewidth} 
       \includegraphics[width=0.8\linewidth]{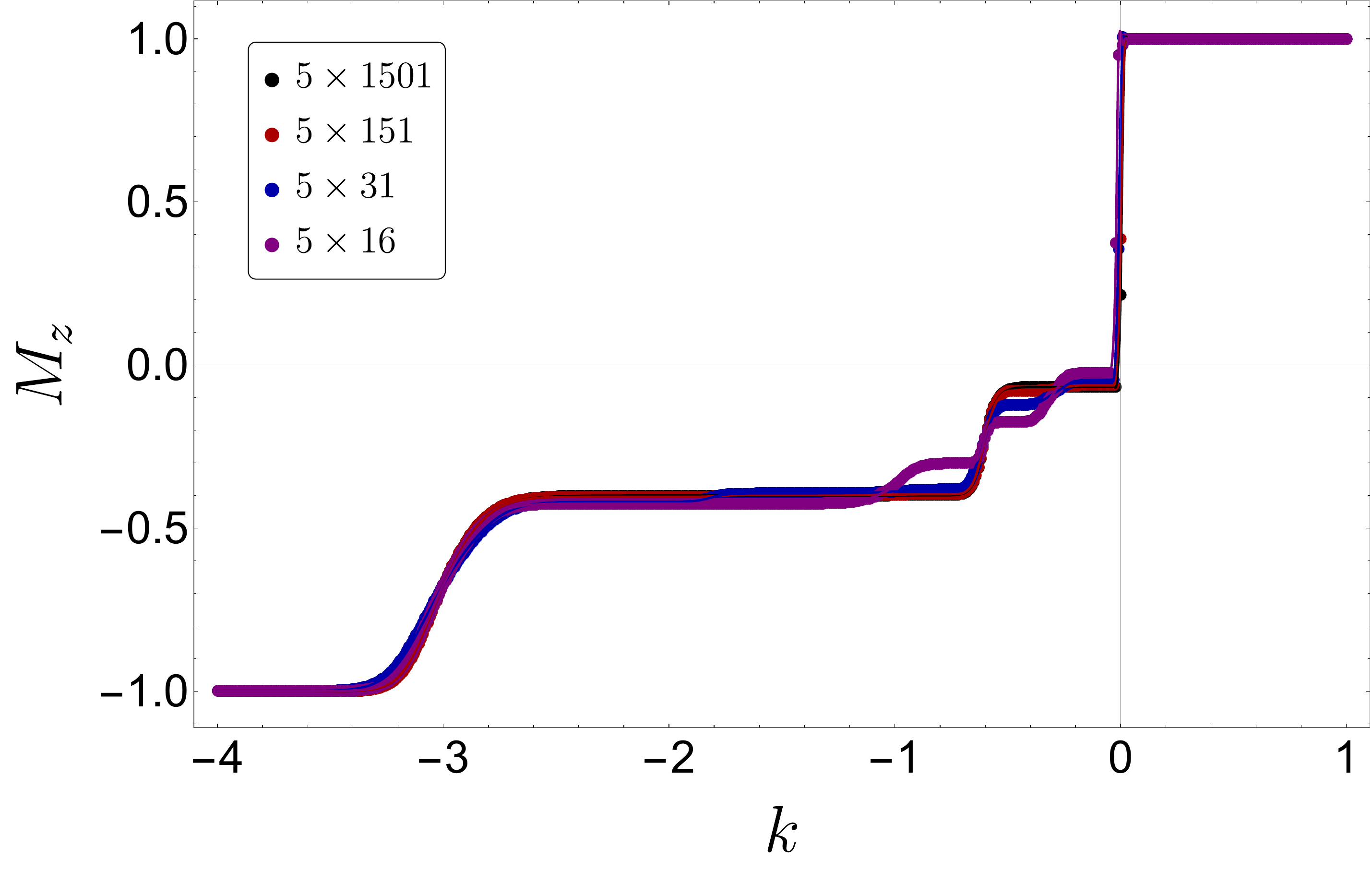}
    \end{subfigure}
    \caption{Magnetization plateaus in the energy minima of the TPMz for negative longitudinal field. (a) Magnetization as a function of $k$ for various sizes (using $e^{-J/T} = 10^{-6}$ in the row-to-row transfer matrix, see text) showing the appearance of plateaus for $k \in [-3J, 0]$. (b) Same for increasing $N = L \times M$ for fixed $L=5$, showing the instability of the highest magnetization plateaus (see this section and Appx.~\ref{appendix:all-models}).}
    \label{fig:plateaus}
\end{figure}

\begin{figure*}[t]
   \includegraphics[width=\textwidth]{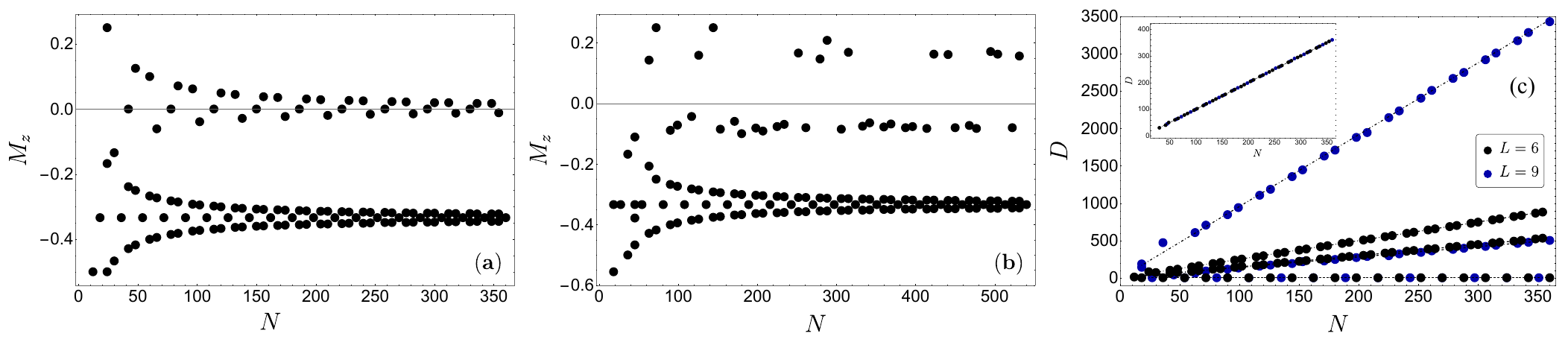}
    \caption{(a) Values of the magnetization at the plateaus in the regime $k \in [-3J, 0]$ as a function of size $N=L \times M$ for fixed $L=6$. (b) Same for $L=9$.
   (c) Degeneracies of the plateaus in (a,b) with $M_z < -0.2$. Inset: same for $M_z >-0.2$. 
    }
    \label{fig:plateaus6and9}
\end{figure*}

\section{Classical Frustrated Phases in the qTPMz}{\label{classical_frustration}}

\subsection{Introduction}

In this section we study the effects of frustration on the energy minima of the TPMz, whose energy function corresponds to \er{QTPMz} at $h=0$,
\begin{equation}{\label{TPMz}}
   \opcatz{E} =  - J \sum_{\{i, j, k\} \in \triangledown} s_i s_j s_k 
   - k \sum_i s_i . 
\end{equation}
Specifically, we find an intricate pattern of {\em magnetization plateaus} corresponding to different classical ground states, emerging due to the competition between the interaction and longitudinal field terms in \er{TPMz}. In order to do this we first review basic ideas about frustrated magnetism and magnetization plateaus. 

After the early works of Wannier and others \cite{wannier1950antiferromagnetism,yoshimori1959a-new-type,elliott1961phenomenological,kaplan1961some}, and further foundational contributions by Toulouse \cite{vannimenus1977theory} and Villain \cite{villain1977spin}, the field of frustrated magnetism has experienced vast advancements; see 
\cite{2011introduction,collins1997review/synthese:,sadoc1999geometrical,2004quantum,diep2020frustrated,moessner2001magnets,starykh2015unusual}
and references therein. Based on the Lieb-Schultz-Mattis theorem \cite{lieb1961two-soluble,affleck1986a-proof}, it was shown by Oshikawa \cite{oshikawa1997magnetization} that magnetization plateaus at zero temperature satisfy the relation
\begin{equation}{\label{magn_plateaus_equation}}
    sV(1 - M_z) \in \mathbb{Z},
\end{equation}
where $s \in \mathbb{Z}_2$ for our work, $V$ defines the translational unit cell which in general might be different in size from the unit cell of the Hamiltonian, $V_0$, and $M_z$ the magnetization of the given plateau. When Eq.~\eqref{magn_plateaus_equation} is not satisfied for $V_0$ then the translational symmetry is spontaneously broken and the translational unit cell is bigger than the unit cell of the Hamiltonian. The above constitutes a theorem only in 1D, despite efforts to extend it to higher dimensions \cite{oshikawa2003insulator}, since for these higher dimensional cases a given degeneracy can also be of topological origin. If translations are spontaneously broken, then Eq.~\ref{magn_plateaus_equation} holds.

The origin of the magnetization plateaus can be classical, when all spins are constrained on $\pm 1$ states as for frustrated models of Ising type \cite{moessner2000two-dimensional,han2008geometric,narasimhan2023simulating,brahma2023emergent}, or quantum, as in Heisenberg models on different frustrated lattices \cite{totsuka1998magnetization,nishimoto2013controlling,sakai2023translational,almeida2023quantum,suetsugu2023emergent}. At the same time, the magnetization might not exhibit a jump upon entering or leaving the given plateau phase \cite{2011introduction}, a phenomenon that as we show below is not encountered in our case. All the classical magnetization plateaus we find appear to be connected by first-order phase transitions. Real materials where the above behaviors can be found extend from classical to quantum magnets, with plateaus both obeying and spontaneously breaking the translational symmetry \cite{shiramura1998magnetization,maignan2004quantum,kikuchi2005experimental,hardy2006magnetism,hase20061/3-magnetization,zhao2010magnetization,wang2011structure,yao20121/3-magnetization,kermarrec2021classical,shangguan2023a-one-third,jeon2024one-ninth}, see Ref.~\cite{yoshida2022frustrated} for a review.

\subsection{Magnetization plateaus in the TPMz}

We now examine the applicability of the above theory for the classical TPM in a field (TPMz). We perform simulations using row-to-row transfer matrices \cite{baxter2016exactly,book_Huang}. We study system sizes $N = L \times M$ with PBC, fixing $L$ and varying $M$. Different $L$'s require different temperatures used to sufficiently approach the zero-temperature ground state, but in most cases a temperature of the order $\exp(-J/T) \approx 10^{-3}$ is enough for convergence. Figure~\ref{fig:plateaus}(a) shows an example of the longitudinal magnetization $M_z$ of the minima of \er{TPMz} as a function of field strength $k/J$ for various sizes, illustrating the emergence of the magnetization plateau structures of the classical ground states. In Fig.~\ref{fig:plateaus}(b) we show an example of the convergence towards the large $N$ limit for fixed $L=5$. 

All the magnetization plateaus that we find below converge to rational values. While some of these plateaus show a very high $V$, they eventually flow to a well-defined small translational unit cell in the thermodynamic limit, where in all cases we keep $L$ fixed and take $M \rightarrow \infty$. Since high-order classical plateaus are unlikely for antiferromagnetic models with short-range interactions \cite{2011introduction}, in our case the large $V$ is due to strong finite size effects. We have also verified in specific cases that the translational symmetry of these plateaus is indeed spontaneously broken.

For $k/J>0$ the unique common ground state of both terms in \er{TPMz} is the ferromagnetic (FM) all-up state, thus giving rise to a classical ground state phase with $M_z=1$. In turn, for $k<-3J$, the energy is dominated by the field term, and the FM all-down state with $M_z = -1$ becomes the unique classical ground state, see Fig.~\ref{fig:plateaus}. 
The situation for $k \in [-3J, 0]$ is more interesting as neither the interaction term nor the  longitudinal field in \er{TPMz} fully dominate, and we now focus on this region where the nontrivial effects of frustration emerge. 

In Fig.~\ref{fig:plateaus6and9}(a,b) we show for $L = 6$ and $L = 9$ and for $k \in [-3J, 0]$ how the magnetization of the plateaus changes as $N$ grows.
The structure of the plateaus shows similar features for both values of $L$. For $M_z \lesssim -0.2$ both system size sequences obey the following rule: for $M=3Q$, with $Q \in \mathbb{Z}$ there is a single plateau at $M_z = -1/3$ for $k \in [-3J, 0]$. The proof of this statement is given in Appx.~\ref{appendix:proof}. For all other system sizes which are not divisible by three, the plateaus occur at different rational values. However, for increasing system size we see an approach to $M_z = - 1/3$ at large $N$. This suggests that in the thermodynamic limit all plateaus for $M_z \lesssim -0.2$ converge to the $M_z = -1/3$ one. In contrast, for $M_z \gtrsim -0.2$ the two cases, $L=6$ and $L=9$, differ: for $L=6$ plateaus saturate around $M_z = 0$, while this does not occur for $L=9$.

For both the $L=6$ and $L=9$ cases, while the plateau at $M_z = - 1/3$ remains stable as $N$ increases, our numerics indicates that the other plateaus do not: with increasing system size the extent of the longitudinal field values of those plateaus diminishes and their location shifts towards $k/J=0$. This tendency can already be seen in Fig.~\ref{fig:plateaus}a. The implication is that in the large $N$ limit and for the system sizes studied here there are three stable values of the magnetization $M_z$ of the classical ground states in the presence of the longitudinal field: $M_z=1$ for $k/J>0$, 
$M_z=-1$ for $k<-3J$, and $M_z=-1/3$ in between. Further intuition on the shifting, their specific values and vanishing of the plateaus is provided in Appx.~\ref{appendix:all-models}. 

The shifting of the plateaus towards the $k/J=0$ point increases the effective degeneracy of the model at this point and has consequences for the structure of the classical ground state manifold of the TPM (i.e., at $k/J=0$). For $3P \times 3Q$ sizes the classical ground states consist of (i) the coexistence of the FM state with $M_z = 1$ and the 3 frustrated states from the $k/J < 0$ region and (ii) the exact degeneracy of all those extra states, as given by the CA rule 60. For $M \neq 3Q$ but $Q \rightarrow \infty$ the extra plateaus encountered for $M_z > -0.2$ for finite sizes give an increased effective ground state degeneracy for the TPM, in the same spirit as for $3P \times 3Q$. Based on the CA description the latter case would show no degeneracies. Our arguments show that, in the thermodynamic limit, not only do they show nontrivial degeneracies, but the asymptotic magnetization of these extra states coincides with the magnetization of the extra states of the $3P \times 3Q$ sizes found from the CAs. Our results are expanded to other strip geometries, including the interesting power of two cases in Appx.~\ref{appendix:all-models}.

The magnetization plateaus for $k \in [-3J, 0]$ are indicative of frustration and can have significant degeneracies. Fig.~\ref{fig:plateaus6and9}(c) shows the multiplicities of the minima in the plateaus of Fig.~\ref{fig:plateaus6and9}(a,b) for values of $M_z \lesssim -0.2$: for the system sizes for which the plateau is exactly at $M_z = - 1/3$ there are only three configurations, while for the other sizes which tend to $M_z = - 1/3$ from above and below with system size, 
the degeneracies grow linearly with $M$. The inset of Fig.~\ref{fig:plateaus6and9}(c) shows the degeneracies for the higher plateaus for both $L=6$ and $L=9$ which disappear in the thermodynamic limit, and whose number also grows linearly (subextensively) with size (as expected from standard TPM arguments, cf.\ Ref.~\cite{sfairopoulos2023boundary}).
\begin{figure}[t]
    \centering
    \includegraphics[width=0.99\linewidth]{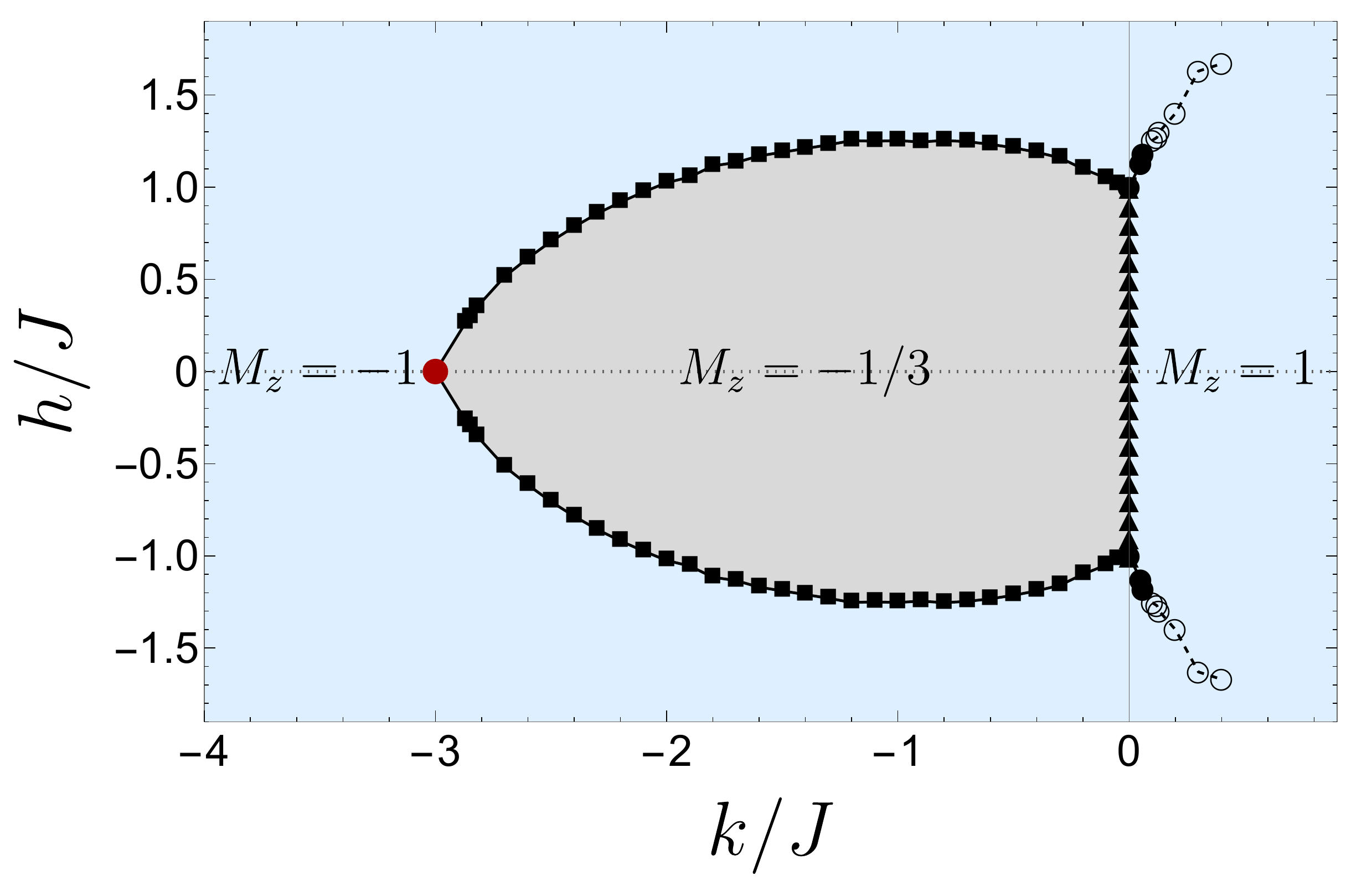}
    \caption{Quantum phase diagram of the qTPMz for sizes $N=3P\times 3Q$. The symbols indicate the location of the phase transitions according to our numerical MPS calculations. The filled circles indicate a first-order transition between the positive FM phase at $k/J>0$ and the PM phase at large $|h|/J$; this transition ends and becomes a crossover, indicated by the empty circles and the dotted line.
    The triangles indicate a first order transition between the frustrated phase at $k/J<0$ and the FM phase at $k/J>0$. 
    The squares indicate the first order transition delimiting the classical frustrated phase.}
     \label{fig:QuantumPhaseDiagram}
\end{figure}

\section{Ground state phase diagram of the qTPMz}{\label{QPT}}

We now study the ground state of the qTPMz, Eq.~\eqref{QTPMz}, as a function of both transverse and longitudinal fields, building on the results of Sec.~\ref{classical_frustration} for the classical TPMz. We use both exact diagonalization \cite{sandvik2010computational} and matrix product state (MPS) techniques with bond dimensions up to 1500 \cite{white1992density,schollwock2011the-density-matrix,fishman2022the-itensor,fishman2022the-itensor2}. Although MPS simulations are usually designed to be used in 1D systems, they can also be applied to higher dimensions \cite{stoudenmire2012studying}, 
as for our 2D models here. In this case, we use a ``snake'' MPS, which although not having a competitive scaling in bond dimension in comparison to its 1D counterpart, it is expressive enough for the study of the qTPMz. For convenience we consider sizes $N = 3P \times 3Q$, with $P,Q \in \mathbb{Z}$, and then extend our results to other system sizes. 

Fig.~\ref{fig:QuantumPhaseDiagram} shows the phase diagram of the qTPMz. The $k/J=0$ line corresponds to the qTPM studied in Ref.~\cite{sfairopoulos2023boundary}: as a function of the transverse field $h/J$ the qTPM hosts a classical phase with the ground states as given by the CA rule 60 for small $|h|/J$, and a quantum paramagnet (QPM) at large $|h|$ with the two phases separated by a first-order phase transition at the self-dual point $|h|/J=1$ \cite{yoshida2014quantum,devakul2019classifying,vasiloiu2020trajectory} 
(with the possible addition of spontaneous breaking of the nontrivial symmetries of the TPM). 

Given the richness of the classical minima of the qTPMz, cf.\ Sec.~\ref{classical_frustration}, it is no surprise that expanding for $k/J \neq 0$ shows a wealth of behaviors in its quantum phase diagram. For $k/J \gtrsim 0$ and starting at small $|h|/J$, the system is in the classical FM phase with $M_z = 1$. As $|h|/J$ is increased, eventually at some $|h|/J \gtrsim 1$ there is a first-order transition to the QPM phase, indicated by the filled circles in Fig.~\ref{fig:QuantumPhaseDiagram}. The transition point grows with increasing $k/J$, and eventually dissolves into a smooth crossover, indicated by the empty circles in Fig.~\ref{fig:QuantumPhaseDiagram}. While the change from a first-order transition to a smooth crossover is clear from the numerics, precisely locating the critical point of the first-order line is much more difficult. However, we expect it to occur for $k/J \lesssim 0.1$. This phase transition is not accompanied by any symmetry breaking.

For $k/J<0$ and small $|h|/J$ the system is in the classically frustrated phase with the three translationally symmetric ground states. By increasing the transverse field this frustrated phase transitions into the QPM via a first-order transition, indicated by the square symbols in Fig.~\ref{fig:QuantumPhaseDiagram}. The first-order transition is accompanied by the spontaneous breaking of the translational symmetry in the classical phase. By decreasing the longitudinal field, the extent of the classical frustrated phase decreases, and eventually terminates at $k=-3J$. At $k \ll -3J$ the system exhibits no phase transitions in any local observables and resembles the properties of a single spin in a transverse field. 

Along the $k/J$ axis there is also a line of first-order transitions at $k/J=0$ for $|h|/J < 1$, see triangular symbols in Fig.~\ref{fig:QuantumPhaseDiagram}. This means that the $|h|/J < 1$ ground states of the qTPM ($k/J=0$) correspond to the coexistence of the positive FM and the frustrated phases of the qTPMz, together with the accidental degeneracies occurring at this point. By implication, the $h/J=1$ self-dual point of the qTPM would correspond to a triple point of the qTPMz where the classical frustrated, FM and QPM phases meet. 

The phase diagram of Fig.~\ref{fig:QuantumPhaseDiagram} is the one we conjecture for the qTPMz in the thermodynamic limit. We base this on our numerics for a sequence of sizes $N$ for systems with $N = 3P \times 3Q$ and PBC. For these sizes, the $M_z = -1/3$ plateau corresponds to the existence of the three classical minima for all $N$, see Fig.~\ref{fig:plateaus6and9}.
The quantum phase diagram of the qTPMz is slightly modified for other system sizes we have analyzed, including the $7P \times 7Q$ and the $8P \times 8Q$, but we expect these differences to disappear in the thermodynamic limit and the general structure of Fig.~\ref{fig:QuantumPhaseDiagram} to hold. For both those cases, the frustrated regime includes multiple classical frustrated phases. 
\section{Vicinity of the $k=-3J$ point, effective Rydberg blockade and hard trimers}{\label{k=-3J}}

\subsection{Classical ground states at $k=-3J$}

\begin{figure}[t]
    \centering
    \begin{subfigure}[b]{0.13\columnwidth}
        \subcaption{}
        \includegraphics[width=\linewidth]{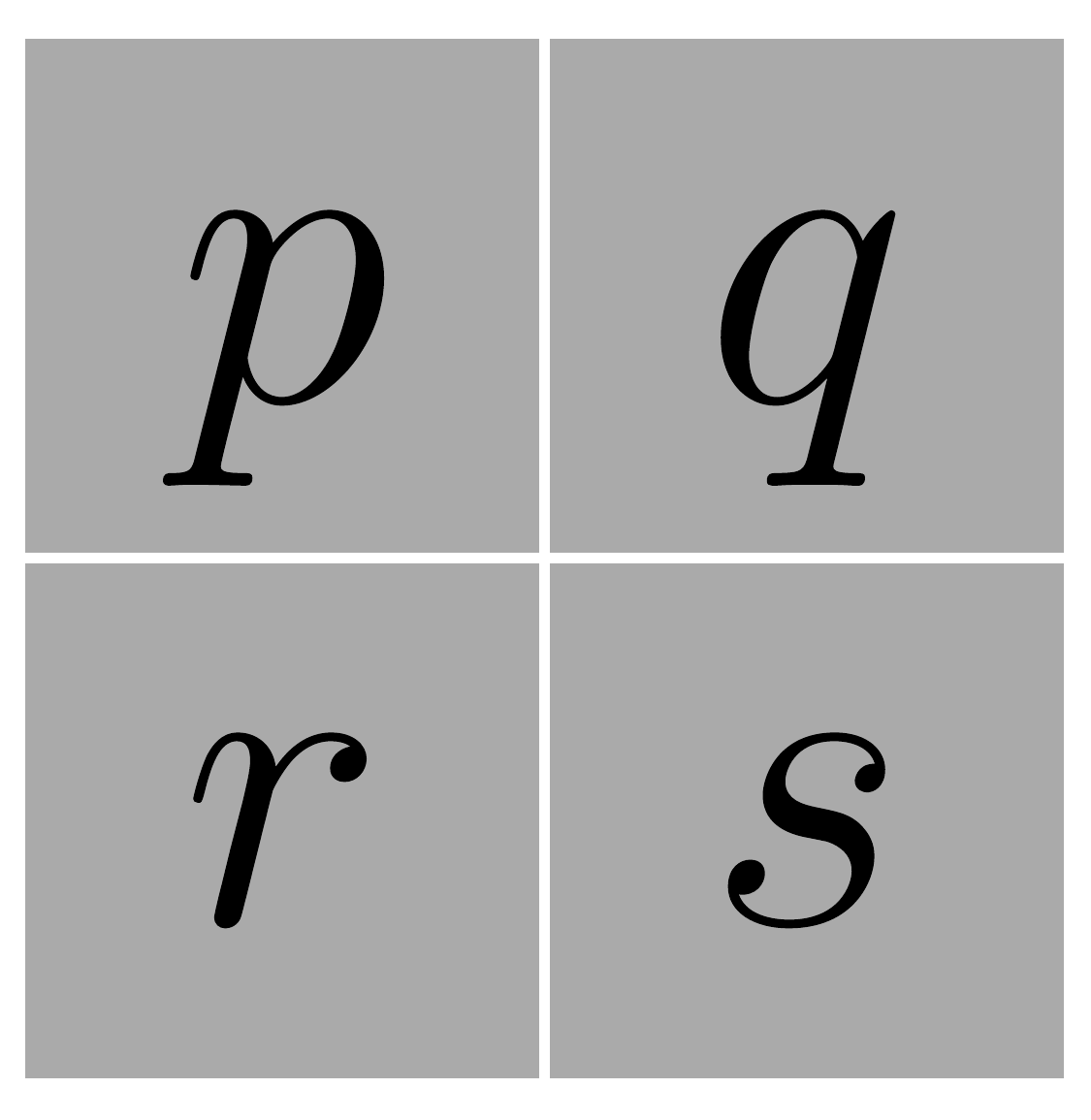}
     \end{subfigure}
     \begin{subfigure}[b]{0.13\columnwidth}
        \subcaption{}
        \includegraphics[width=\linewidth]{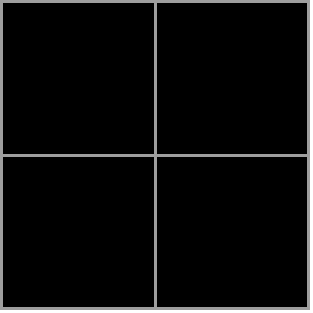}
     \end{subfigure}
     \begin{subfigure}[b]{0.13\columnwidth}
        \subcaption{}
        \includegraphics[width=\linewidth]{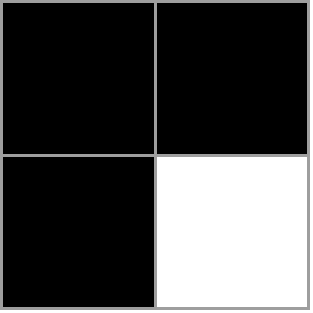}
     \end{subfigure}
     \begin{subfigure}[b]{0.13\columnwidth}
        \subcaption{}
        \includegraphics[width=\linewidth]{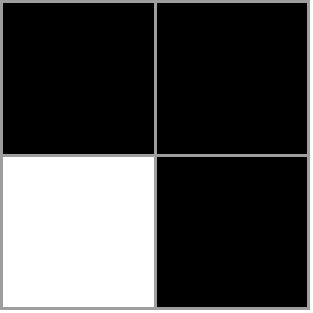}
     \end{subfigure}
     \begin{subfigure}[b]{0.13\columnwidth}
        \subcaption{}
        \includegraphics[width=\linewidth]{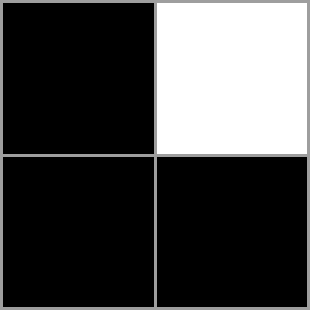}
      \end{subfigure}
      \begin{subfigure}[b]{0.13\columnwidth}
         \subcaption{}
         \includegraphics[width=\linewidth]{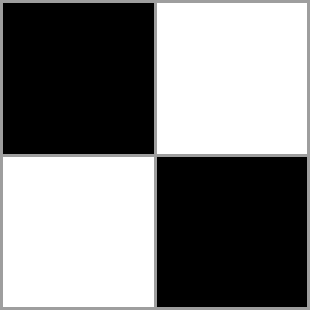}
      \end{subfigure}
     \begin{subfigure}[b]{0.13\columnwidth}
        \subcaption{}
        \includegraphics[width=\linewidth]{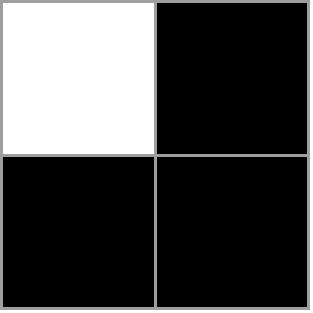}
     \end{subfigure}
     \begin{subfigure}[b]{0.23\columnwidth}
      \includegraphics[width=\linewidth]{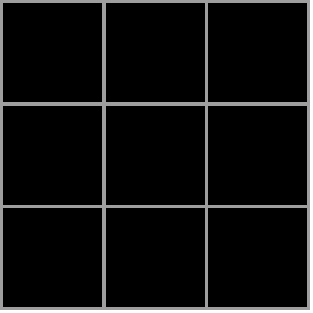}
      \subcaption{}
   \end{subfigure}
   \begin{subfigure}[b]{0.23\columnwidth}
      \includegraphics[width=\linewidth]{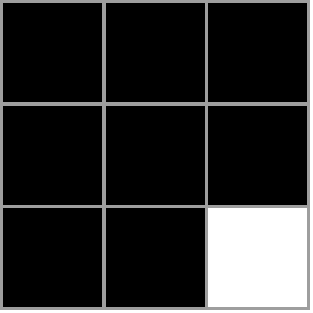}
      \subcaption{}
   \end{subfigure}
   \begin{subfigure}[b]{0.23\columnwidth}
      \includegraphics[width=\linewidth]{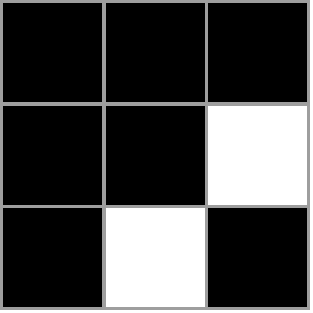}
      \subcaption{}
   \end{subfigure}
   \begin{subfigure}[b]{0.23\columnwidth}
      \includegraphics[width=\linewidth]{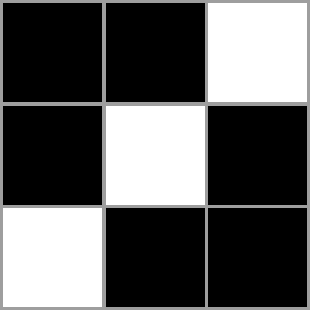}
      \subcaption{}
   \end{subfigure}
     \caption{(a) Labelling of sites in \era{eq:rule60}{eq:km3}. (b-g) Local ground state tiles for the classical minima of the TPMz at $k=-3J$, where white/black indicate up/down spin.
     (h-k) Classical ground states of the TPMz at $k=-3J$ for size $N=3\times3$ with PBC, obtained from tiling space with tiles (b-g) (up to configurations equivalent by translations).}
     \label{fig:tiles}
\end{figure}

The classical TPMz has two special points where finding the energy minima of \er{TPMz} can be formulated as a constrained satisfaction problem in $\mathbb{F}_2$. One is the TPM limit, $k=0$, studied in Ref.~\cite{sfairopoulos2023boundary}, whose minimum energy configurations coincide with the cycles of the CA rule 60 and its evolution is described by the local rule
\begin{equation}
   s = p + q \mod 2, 
   \label{eq:rule60}
\end{equation}
where $\{p,q,s\} $ take values in $\mathbb{F}_2$ and correspond to binary variables in a two-dimensional square neighbourhood as shown in Fig.~\ref{fig:tiles}(a). Since the triangular lattice, cf. Fig.~\ref{fig:TPM_lattice}, can be deformed into a square lattice by shearing, obeying the local rule Eq.~\eqref{eq:rule60} is equivalent to minimizing the interaction in \er{TPMclassical} (where $\sigma_i \leftrightarrow 1 - 2 p$, $\sigma_j \leftrightarrow 1 - 2 q$, etc), and the minima of the TPM can be constructed propagating \er{eq:rule60} row 
to row \cite{sfairopoulos2023boundary}.

\begin{figure}[t]
   \centering
   \includegraphics[width=0.8\linewidth]{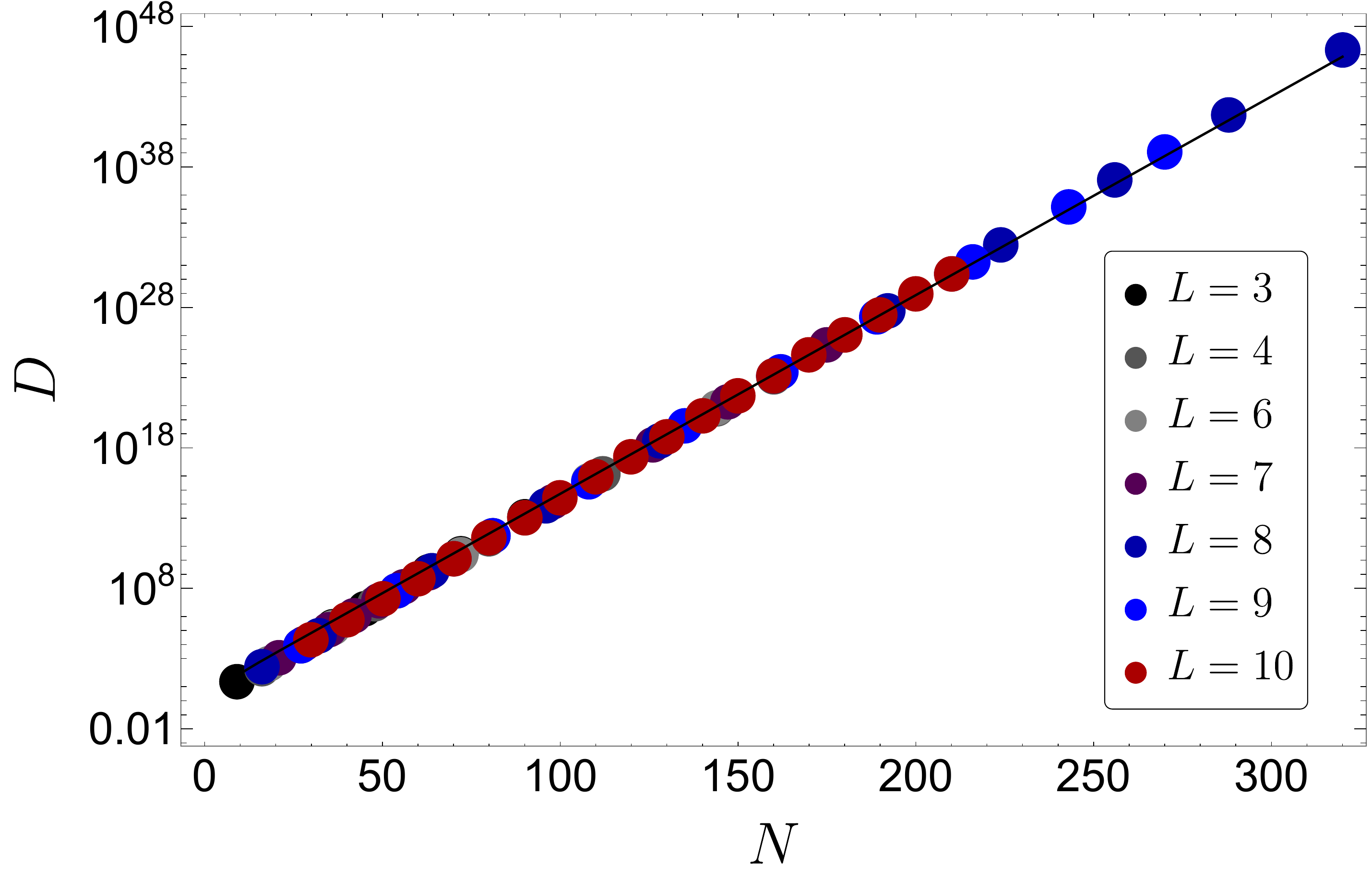}
   \caption{Degeneracy of the minima of the TPMz at $k=-3J$ as a function of size $N$ for several fixed $L$.}
    \label{fig:Degeneracy_scaling_at_PhTr}
\end{figure}

The second special point is that of \er{TPMz} at $k = -3J$. For this value of the longitudinal field, the minima of \er{TPMz} satisfy the constraint 
\begin{equation}
   pq + qs + sp = 1 \mod 2 , 
   \label{eq:km3}
\end{equation}
in the convention of Fig.~\ref{fig:tiles}(a). Note that in contrast to \er{eq:rule60}, the classical ground states of \er{TPMz} at $k = -3J$ cannot be generated by evolving a one-dimensional CA. Instead, they are formed by {\em tiling}, where the allowed tiles are those shown in Fig.~\ref{fig:tiles}(b-g), where a black site denotes state 1 (flipped spin) and the white site state 0 (up spin). As an example, in Fig.~\ref{fig:tiles}(h-k) we show the ground states of a $N=3\times3$ system with PBC obtained by combining the tiles of Fig.~\ref{fig:tiles}(b-g): for this system size a total of 22 ground states can be accommodated [those of Fig.~\ref{fig:tiles}(h-k) plus their translations]. Generalizing to other sizes, one can calculate the classical ground state degeneracy via transfer matrix techniques: as shown in Fig.~\ref{fig:Degeneracy_scaling_at_PhTr}, at the $k=-3J$ point, the TPMz has an exponential number of ground states, with clear evidence for $1/3$ of states minimizing the energy (in contrast to the algebraic degeneracy at other values of $k/J$, cf.\ Fig.~\ref{fig:plateaus6and9}).

\subsection{Effective description in terms of a Rydberg blockade}

As we will now show, the properties of the TPMz at $k=-3J$ lead to an effective description of the qTPMz around that point (that is, for values of the longitudinal and transverse field strengths $|k + 3J| \ll J$ and $h \ll J$) in terms of an effective model with a two-dimensional ``Rydberg-blockade'', in analogy with the much studied PXP model for one-dimensional Rydbergs \cite{fendley2004competing,lesanovsky2011many-body,lesanovsky2012interacting,turner2018weak}. 

\begin{figure}
   \centering
   \includegraphics[width=0.7\linewidth]{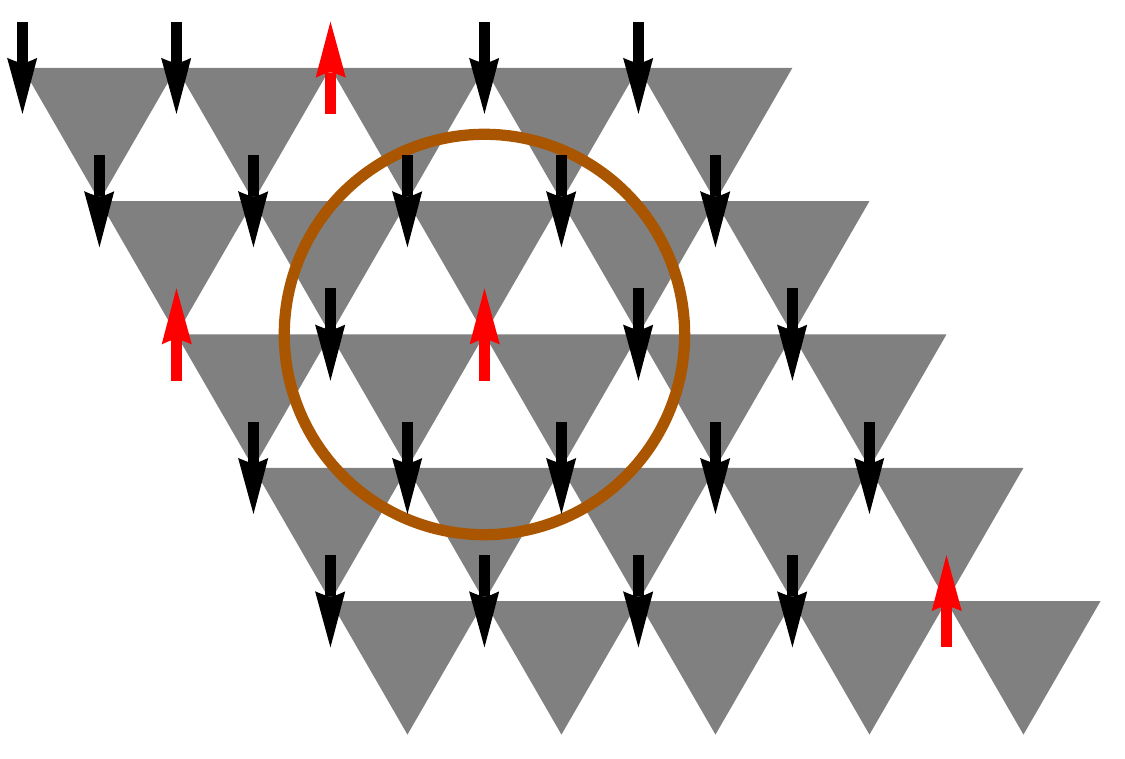}
   \caption{A configuration satisfying the Rydberg blockade condition with $m_\triangledown=0$ and PBC, with excited (up) spins (in red) in a background of down spins (in black).}
    \label{fig:Rydbergconnection}
\end{figure}

As we aim to study the behavior of the qTPMz near the $k=-3J$ point, we write the Hamiltonian Eq.~\eqref{QTPMz} as 
\begin{align}
   \opcatqz{H}  = 
      + 2J & \sum_{\{i, j, k\} \in \triangledown}^N 
         \text{CZ}_{ij} \text{CZ}_{jk} \text{CZ}_{ki}
      \nonumber \\
      - h & 
      \sum_{i}^N X_{i} 
      - (k+3J) \sum_i^N Z_i, 
   {\label{QTPMzv2}}
\end{align}
where we've used the fact that $\frac{1}{2}(Z_i + Z_j + Z_k - Z_i Z_j Z_k ) = \text{CZ}_{ij} \text{CZ}_{jk} \text{CZ}_{ki}$.
If we define a new defect variable at a plaquette, $\triangledown$, formed by spins at sites $(i,j,k)$ as
\begin{equation}
   m_\triangledown  = 
      1- \frac{1}{2} \left( Z_i Z_j Z_k - Z_i - Z_j - Z_k \right) ,
   {\label{eq:m}}
\end{equation}
we can see that the minima of \er{QTPMzv2} at $k = -3J, h=0$ are configurations for which $m_\triangledown=0$ for all $\triangledown$, while plaquettes with $m_\triangledown=1$ represent excitations away from the set of these minima. We can then write \er{QTPMzv2} as
\begin{equation}
   \opcatqz{H}  = 
      2 J \sum_{\triangledown}^N m_\triangledown 
      - h \sum_{i}^N X_{i} 
      - (k+3J) \sum_i^N Z_i 
      + {\rm const.}  
   {\label{QTPMzv3}}   
\end{equation}
When $|h|, |k+3J| \ll J$, this is reminiscent of the Rydberg problem described above, as excitations away from the $\{ m_\triangledown = 0 \}$ subspace are highly suppressed. Specifically, to require that all $m_\triangledown = 0$ is equivalent to the nearest neighbour blockade condition for up spins in the triangular lattice, 
see Fig.~\ref{fig:Rydbergconnection}. This means that just as the 1D Rydberg case leads to the PXP model, the dynamics under $\opcatqz{H}$ for $|h|/J$ and $|k+3J|/J$ small is effectively described by (cf.\ Rydbergs on a square lattice \cite{ji2011two-dimensional})
\begin{equation}{\label{p6x}}
   \opcatpppxppp{H} = - h \sum_{i}^N P_{6}^{(i)} X_{i} - \delta k  \sum_{i}^N Z_{i},
\end{equation}
where we have dropped the constant term, $\delta k$ denotes the differential of $k$ and $P_{6}^{(i)}$ projects onto the nearest neighbors, $j$, of $i$ to be all pointing down, 
\begin{equation}
   P_{6}^{(i)} = \bigotimes_{\langle i,j \rangle}^6 \frac{1}{2} \left( 1 - Z_{j} \right) .
   \label{eq:P6}
\end{equation}

\begin{figure}[t]
   \centering
   \includegraphics[width=0.7\linewidth,trim={20mm 0 0 0},clip]{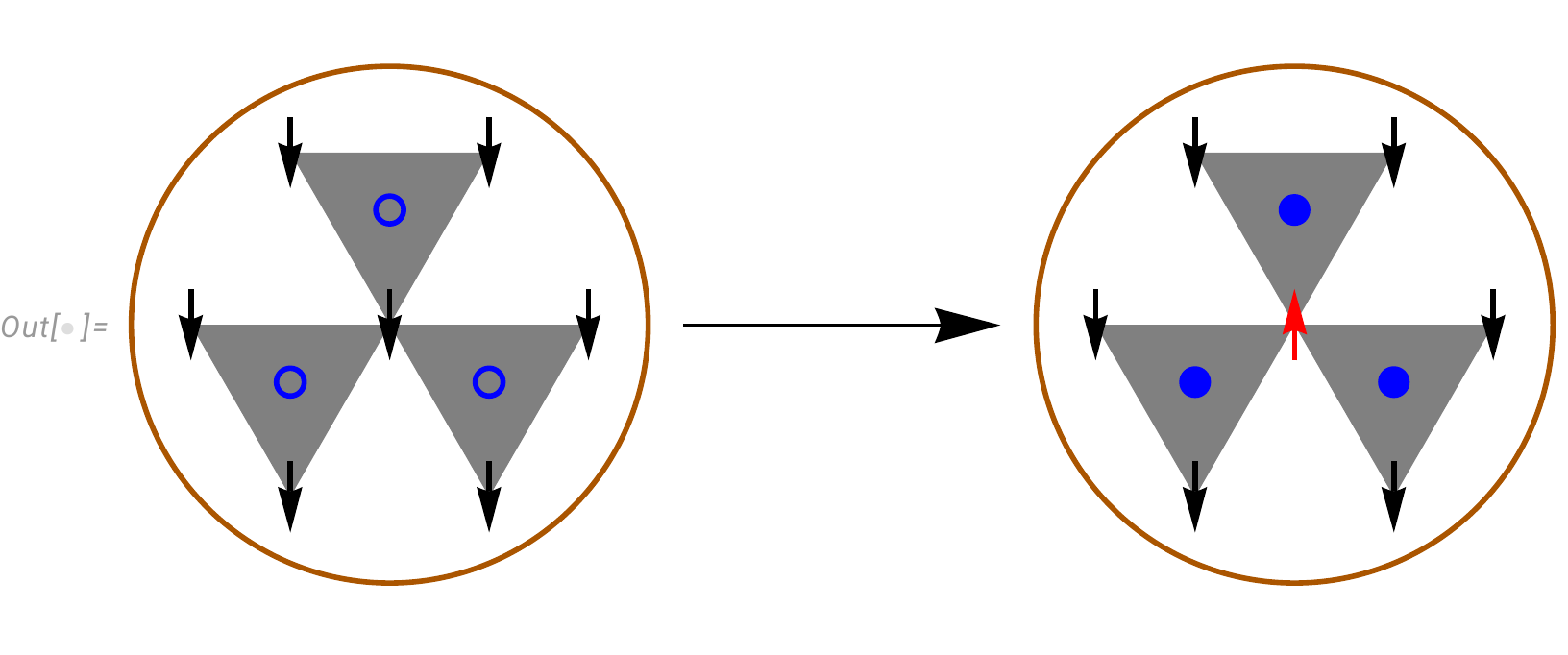}
   \caption{Effective constraint of the qTPMz in the vicinity of the $k=-3J,h=0$ point expressed in terms of composite bosons. Fermions occupy the sites of the dual lattice. Flipping a spin is accompanied by the creation of three fermions.}
    \label{fig:compositebosons}
\end{figure}

\subsection{Lattice gauge theory and hard trimers}{\label{LGTs}}

The constraints in the PXP model Eq.~\eqref{PXP} that embody the Rydberg blockade for a 1D lattice can be expressed with the help of auxiliary degrees of freedom, giving rise to locally conserved quantities and an effective lattice gauge theory (LGT). This connection was studied in Refs.~\cite{surace2020lattice,pan2022composite,cheng2022tunable,cheng2023variational,cheng2023gauge,cheng2024emergent}, establishing the significance of Rydberg atom systems as cold-atom simulators of gauge theories \cite{halimeh2023cold-atom}. 

We can apply a similar approach to that of Ref.~\cite{surace2020lattice} here. We express the constraint in \er{p6x} that no nearest neighbouring spins can be in the down state with the help of auxiliary fermions: we place the fermions on the dual lattice (i.e., the plaquettes), with creation and annihilation operators $f_{\triangledown}^\dagger$ and $f_{\triangledown}$ for plaquette $\triangledown$. We stipulate that flipping a down spin implies creating fermions in its neighbouring plaquettes, see Fig.~\ref{fig:compositebosons}. The corresponding LGT takes the form 
\begin{equation}
   \opcatlgt{H} = 
   - h \sum_{i}^N \left(
   S^{+}_{i} 
   f_{\triangledown_{i,1}}^\dagger
   f_{\triangledown_{i,2}}^\dagger
   f_{\triangledown_{i,3}}^\dagger
   + {\rm h.c.} 
   \right)
   - \delta k  \sum_{i}^N Z_{i} ,
   \label{eq:LGT}
\end{equation}
where $\triangledown_{i,1}, \triangledown_{i,2}, \triangledown_{i,3}$ are the three plaquettes around site $i$. The redundancy in the degrees of freedom is evident from the gauge symmetry of \er{eq:LGT} with generators
\begin{equation}
   G_\triangledown \equiv f_\triangledown^\dagger f_\triangledown - \frac{1}{2} \sum_{i \in \triangledown}^N \left( Z_i + 1 \right) ,   
\end{equation}
where $i \in \triangledown$ indicates the sites that define the plaquette $\triangledown$. Fixing the gauge condition $G_\triangledown = 0$ for all plaquettes provides the equivalence with the ``Rydberg'' description of the qTPMz near the $k=-3J$ point, \er{p6x}. This LGT is similar in appearance to the mean-field theory defined for quantum spin liquids \cite{book_Wen,book_Sachdev}.

\begin{figure}[t]
   \centering
   \includegraphics[width=0.7\linewidth]{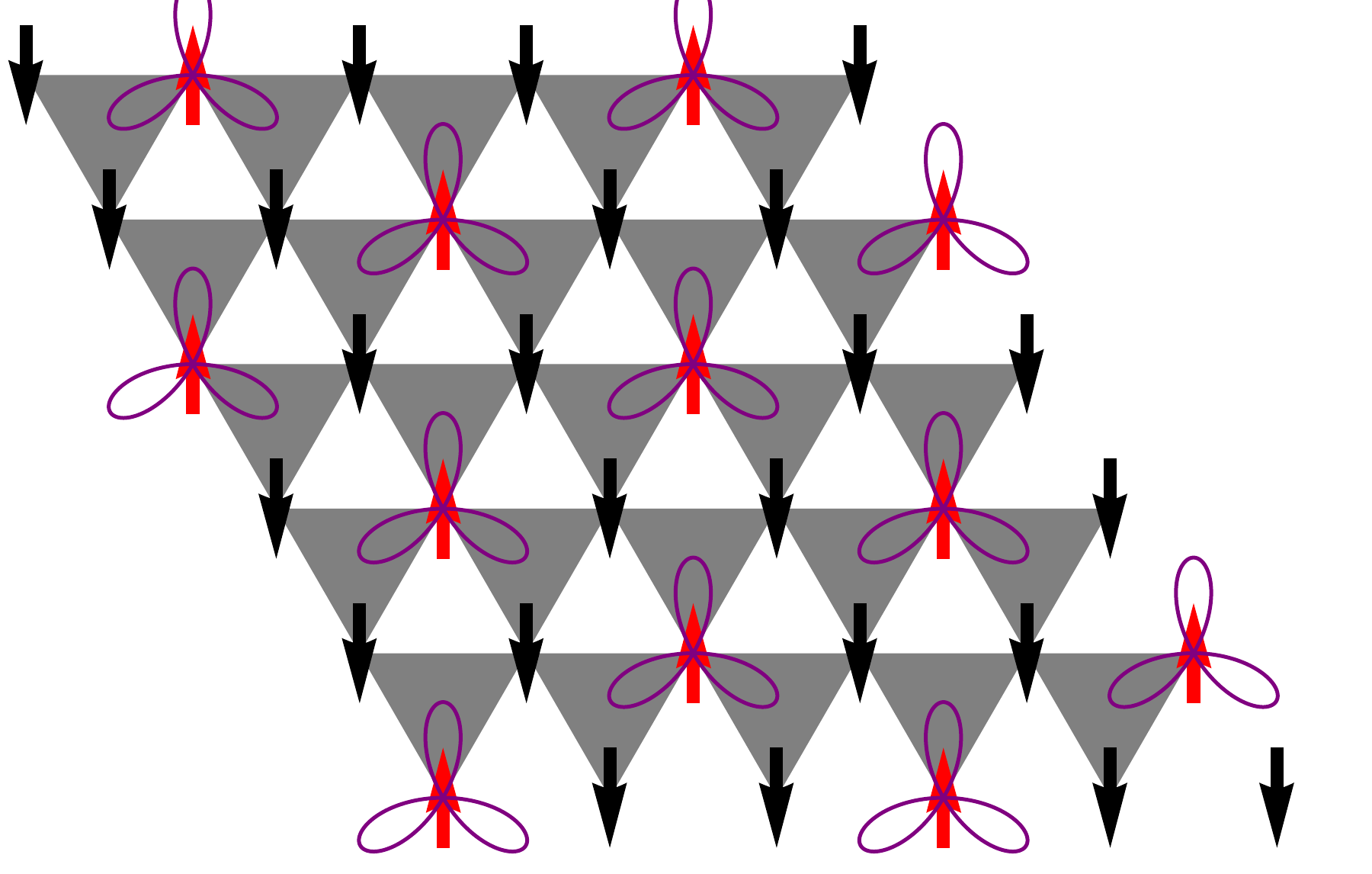}
   \caption{Blockade condition in terms of hard trimers (in purple) centred on the excited spins (in red). 
   We show the example of maximum trimer occupancy corresponding to an ordered arrangement of trimers in every third site. There exist three such translationally symmetric configurations.
   }
    \label{fig:trimers}
\end{figure}

As for quantum dimers \cite{moessner2001ising,moessner2001resonating,moessner2001short-ranged,misguich2002quantum,ralko2005zero-temperature,yan2021topological,yan2022triangular}, we can connect the LGT formulation of our problem to one in terms of quantum {\em trimers}. Consider placing a trimer on each site with an up spin. If we impose that a plaquette can only have one trimer ``arm'' on it, a configuration of hard trimers corresponds to a blockaded configuration of the spins; see Fig.~\ref{fig:trimers} for an illustration. This means that the set of all spin configurations that obeys the ``Rydberg'' constraints is the same as the set of all configurations with any number of trimers, i.e., a mixture of trimers and monomers, meaning sites without trimers. Lastly, note that, contrary to other works in the literature where trimers can occur in various shapes \cite{jandura2020quantum,iskin2022dimers,giudice2022trimer,cheng2022fractional}, here we are constrained by geometry to a single one.

In the trimer description the classical ground states of the TPMz around $k=-3J$ are as follows. At $k \gtrsim -3J$, the minima are configurations that maximize the number of trimers; for sizes such as $N=3P\times3Q$ this is achieved by perfect (i.e., fully packed) and ordered tilings such as the one shown in Fig.~\ref{fig:trimers}, with $M_z = -1/3$ and degeneracy of three, while for other sizes they are dense but not perfect trimer arrangements and their multiplicity is generally exponential, cf.\ Fig.~\ref{fig:plateaus6and9} and Appx.~\ref{appendix:all-models}. For $k \lesssim -3J$, in contrast, there is a unique FM minimum with no trimers and $M_z = -1$. At the point $k = -3J$ all monomer-trimer arrangements are equivalent giving rise to the exponential degeneracy of the minima, cf.\ Fig.~\ref{fig:Degeneracy_scaling_at_PhTr}. Note that in contrast to, say, fully packed dimers on the square lattice, see e.g. Refs.~\cite{moessner2011quantum,henley2010the-coulomb,chalker2017spin},
in our case there is no ``Coulomb'' phase at perfect covering.

\section{Variational ground state near $k=-3J$ and candidate topological ordered phase}\label{sec:topo}

Following the LGT construction in Sec.~\ref{LGTs}, we introduce composite spins, $\tau_i$, by combining the spin at site $i$ with its three neighbouring fermions at plaquettes $\triangledown_{i,1}, \triangledown_{i,2}, \triangledown_{i,3}$, cf.\ Ref.~\cite{pan2022composite,cheng2023variational}. The two states of such a composite object will look like $\ket {\Downarrow} \equiv \ket {\downarrow} \otimes \ket {000}$ and $\ket {\Uparrow} \equiv \ket {\uparrow} \otimes \ket {111}$, and correspond to placing a monomer or a trimer in the lattice, respectively. For example, in this notation, the negative FM ground state for $k < -3J$ corresponds to all (composite) spins being $\ket {\Downarrow}$. Note that, since the composite spins include the auxiliary fermions, they do not obey the standard anticommutation relation of Pauli matrices. In terms of the composite spins, the Hamiltonian Eq.~\eqref{eq:LGT} is just the sum of a transverse and a longitudinal field terms, 
\begin{equation}{\label{composite free spins hamiltonian}}
   H_\tau = - h \sum_{i}^N X_{i}^{\tau} + \delta k \sum_{i}^N Z_{i}^{\tau} ,
\end{equation}
where the superscript $\tau$ indicates that these are operators on the composite spins. In this prescription, the Rydberg blockade (or hard trimer condition), cf.\ Fig.~\ref{fig:Rydbergconnection}, is imprinted in the spins via the exclusion principle of the fermions. 

\begin{figure*}
   \includegraphics[width=1.95\columnwidth]{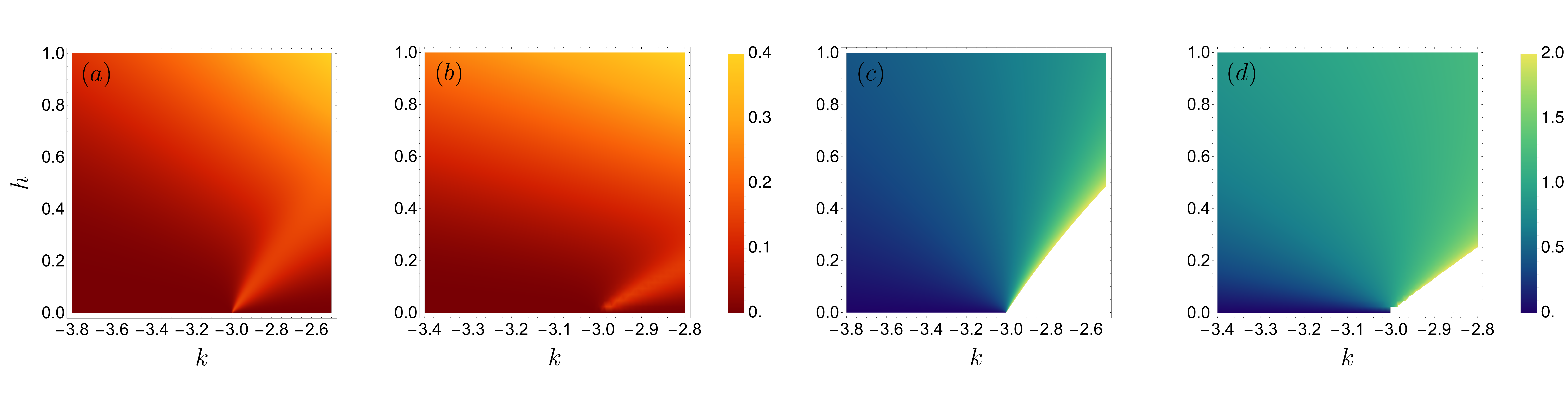}
 \caption{(a) Infidelity of the ansatz Eq.~\eqref{eq:var2} 
w.r.t. the numerically exact ground state of the qTPMz as a function of $k$ and $h$ in the region of the $k=-3J,h=0$ point for $N=3\times7$. 
(b) Same for $N=5 \times 5$. (c) Optimal variational parameter $v$ for the case of panel (a) for $u=1$. (d) Same for $N=5 \times 5$. The region of the phase space coloured white indicates values of $v$ with much larger values than the colour scale.
}
 \label{fig:BCSwavefunction}
\end{figure*}

We first consider the FM region at large negative longitudinal field, see Fig.~\ref{fig:QuantumPhaseDiagram}, where the ground state has no occupied fermions and all spins pointing down, thus locally being $\Downarrow$. As the absolute value of the field strength is decreased towards $k=-3J$ we expect the ground state to hybridize with those states where (composite) spins are flipped up. On the other limiting case of $k > -3J$ the ground state will have all \footnote{or nearly all, if the system sizes are not of the form $N = 3P \times 3Q$. In general, this is a tiling problem.} its fermions occupied. Let us approximate the ground state of the system with the state (up to a normalization factor)
\begin{align}
   \begin{split}
   \ket {\Psi} 
   &
   \approx \prod_{i}^N \left( u  + v \, S^{\tau,+}_i \right) \ket{\text{vac}} \\
   &
   \approx \sum_{\Lambda} u^{N_{\Downarrow}(\Lambda)} v^{N_{\Uparrow}(\Lambda)} \ket {\Lambda},
   \label{eq:var2}
   \end{split}
\end{align}
with $S^{\tau,+}_i \equiv S^{+}_{i} f_{\triangledown_{i,1}}^\dagger f_{\triangledown_{i,2}}^\dagger f_{\triangledown_{i,3}}^\dagger$ the flip-up operator for a composite spin at site $i$ of the original lattice, $\ket{\text{vac}}$ the vacuum state with $\ket{\text{vac}} = \prod_i^N \ket {\downarrow}_i \otimes \prod_j^N\ket {0}_j$ which is the ground state of the system for $k \ll -3J$, and where $\{ \Lambda \}$ denotes the set of allowed configurations and $N_{\Downarrow, \Uparrow}(\Lambda)$ the number of down or up spins in the configuration $\Lambda$, with $N_{\Downarrow}(\Lambda) + N_{\Uparrow}(\Lambda) = N$. The first line of Eq.~\ref{eq:var2} represents the ground state ansatz for the effective LGT, and the second the ground state only in tern of the spins. 

The ansatz Eq.~\eqref{eq:var2} captures the known limiting behavior: for $v \rightarrow 0$ it recovers the FM ground state of the qTPMz Hamiltonian for $k \ll -3J$, corresponding to the states $\Lambda$ with $N_{\Uparrow}(\Lambda) = 0$; for large $v$ it recovers the maximum covering of trimers, corresponding to the states $\Lambda$ with $N_{\Uparrow}(\Lambda) = N/3$ for $3P \times 3Q$ sizes in the large size limit; and at $k = -3J, h=0$ it is an equal superposition of all trimer configurations (for $u$ constant). Assuming the form of Eq.~\eqref{eq:var2} captures also the ground state properties for small $h$ and $\delta k$, it is interesting to observe that this BCS-like wavefunction can support $\mathbb{Z}_2$ topological order for intermediate values of $v$ for other lattices \cite{li2007topological,zhou2017quantum,cheng2023variational}. 

We now test the quality of the approximation of Eq.~\eqref{eq:var2}. We compute the ground state, $\ket{\rm g.s.}$, via exact diagonalisation for small systems, and calculate the (in)fidelity of the variational approximation $F = 1 - \Braket{ \Psi | {\rm g.s.} }$. Fig.~\ref{fig:BCSwavefunction}(a-b) shows $F$ for sizes $N=3 \times 7$ and $N=5 \times 5$, respectively, at the optimal value of the variational parameter $v$ at each state point. In particular, for small transverse fields $h//J$, we see that $\ket{\Psi}$ seems to provide an accurate representation of the ground state of the qTPMz, at least for these small sizes where we can make the comparison.  

In Fig.~\ref{fig:BCSwavefunction}(c-d) we show the optimal values of $v$. We see that $v$ tends to vanish for small $h/J$ and $k < -3J$, while $v \gg 1$ for small $h/J$ and $k > -3J$, in accordance with the discussion above. For $|h|/J \gtrsim 0$, we anticipate the existence of a quantum spin liquid regime between the PM and frustrated phases, in analogy to Refs.~\cite{verresen2021prediction,samajdar2021quantum}. The region of intermediate values of $v$ and high fidelity in Fig.~\ref{fig:BCSwavefunction}(c-d) would be the candidate spin liquid regime in our case. More specifically, this regime would be close to the $k=-3J$ point to exploit its exponential ground state degeneracy but on the outside of the classical frustrated phase and on both sides of the horizontal axis, where the ground state is expected to be comprised of a mixture of monomers and trimers. From our numerics, however, we identify a smooth crossover and no signatures of a phase transition between the PM phase and the conjectured spin liquid regime. 

The existence of a spin liquid could be further confirmed by computing the topological entanglement entropy (TEE) \cite{kitaev2006topological,levin2006detecting}: the entanglement entropy of a gapped phase obeying perimeter law scales as $S = aL - \gamma$ (with $L$ the perimeter), with the TEE
$\gamma=\ln{2}$ for a $\mathbb{Z}_2$ spin liquid. We computed $\gamma$ by considering cylindrical geometries and bipartitioning the system \cite{jiang2012identifying,jiang2013accuracy}, but found it to suffer from too strong finite size effects to confirm the existence of a spin liquid phase in the qTPMz (as also occurs in other systems, see e.g. Refs.~\cite{gong2013phase,zhu2013weak,gong2014plaquette,zou2016spurious,williamson2019spurious}). 

In order to argue more firmly in favour of a spin liquid phase, nonlocal order parameters like the ones in Refs.~\cite{verresen2021prediction,semeghini2021probing,cheng2023variational} would need to be formulated and studied. Furthermore, since a spin liquid is characterized by its fractionalized excitations, a study of whether spinon excitations exist would need to be performed. We were unable to establish whether this candidate spin liquid regime constitutes a separate phase to the QPM for further negative $k$: for the MPS numerics all local operators vary smoothly and the gap in the ground state (and the associated correlation length) shows no singular structure. Lastly, structure factors were also unable to indicate signatures of any resonating valence bond solid phase inbetween the QPM and the frustrated phases, as for example in Ref.~\cite{celi2020emerging}. These results leave us with two possibilities: a paramagnetic phase that extends till the $k=-3J$ point, without any intermediate phase or a resonating valence bond phase with a smooth crossover to the paramagnetic phase for smaller longitudinal fields. Either of these scenarios would need to be tested on larger system sizes so that a definite answer is provided.

Experimental and theoretical works that studied Ising systems on the kagome and ruby lattices, for example Refs.~\cite{semeghini2021probing,samajdar2021quantum,verresen2021prediction,narasimhan2023simulating}, found evidence for the existence of a spin liquid phase between nematic and staggered phases, or trivial and valence bond solid phases. If a spin liquid phase exists in the qTPMz, one can argue that this is a more generic situation; the construction of the LGT and the relation to the Rydberg physics is similar to all other $p$-XORSATs, including the quantum rule 150 and the quantum square pyramid model in a longitudinal field, as explained in Appx.~\ref{appendix:XORSATs}.

\section{Discussion and outlook}{\label{conclusions}}

In this work we studied the quantum triangular plaquette model in a longitudinal field (qTPMz). We found that, although previous works have studied the classical TPM in a longitudinal field \cite{sasa2010thermodynamic,garrahan2014transition,turner2015overlap}, there was a lack of understanding of the ground state properties of the model for negative longitudinal field strengths. Our results here show the presence of a variety of classical ``phases'' with distinctive magnetization plateau structures. We use the word ``phases'' in quotation marks because the finite size scaling for these systems is tricky to perform and interpret, see e.g. Ref.~\cite{sfairopoulos2023boundary}. However, we manage to prove that systems with sizes $3P \times 3Q$ with $P, Q$ integers, possess only one classical phase with magnetization $M_z = -1/3$. We prove this statement in Appx.~\ref{appendix:proof}. Furthermore, we provide extra results for other system sizes and their finite size scaling in Appx.~\ref{appendix:all-models}. Based on these classical results, we can infer some general properties of the model: (i) many magnetization plateaus tend to shrink in size for small negative longitudinal fields, $k$, and collide to the $k=0$ point where they give a higher effective ground state degeneracy of the classical TPM phase, (ii) even more, the classical TPM corresponds to the coexistence of a trivial phase and the frustrated phases with the additional ``accidental'' degeneracy of those added states from (i), (iii) moving to the $k=-3J$ point, we construct the nondirectional constraint that gives rise to the exponential ground state manifold. 

Moving on to our results that concern properties of the system that come from the interplay of the TPMz and quantum fluctuations from the transverse field term, our main result is the phase diagram of Fig.~\ref{fig:QuantumPhaseDiagram}. The phase diagram of the qTPMz includes a classical trivial phase for $k/J>0$, $h/J$ small which is continuously connected to the quantum paramagnetic phase through a first-order quantum phase transition which ends at a critical point. This quantum phase transition line continues as a crossover for larger positive $k$ values. On the other hand, for negative $k$ we find a classical frustrated phase (or more phases, if a sequence of system sizes that is not $3P \times 3Q$ with integers $P, Q$ is used) which is connected to the quantum paramagnet with a first-order quantum phase transition. These two phase transition lines meet at $k/J=0$, $h=J$ at the phase transition point of the qTPM \cite{sfairopoulos2023boundary}. Retrospectively, we understand this point as a triple point, where the phase transitions for $k/J<0$, $k/J>0$ and the $k/J=0$ ones meet. Regarding the last one, this transition is not quantum in nature, but distinguishes two classical phases. 

Moving closer to the $k\approx -3J$ region the first-order quantum phase transition line approaches the $h$-axis and ends at the $k=-3J$ point. For $k > -3J$, we observe no signatures of a phase transition and, we argue, that for negative enough values of $k$ the model becomes trivial. Close to the $k=-3J$ point, we argue that the model can be equally described as a lattice gauge theory, as a model of quantum trimers, or as a blockaded Rydberg model. At the same time, we find a variational wavefunction which approximates accurately enough the ground state and bears similarities to a resonating valence bond solid wavefunction \cite{auerbach_interacting_1994,zhou2017quantum}. In our view this is a significant result as it combines the physics of quantum dimers/trimers, of lattice gauge theories, of Rydberg models with frustration and of models of glasses and dynamical facilitation. Lastly, we find no clear signatures for a $\mathbb{Z}_2$ topologically ordered phase for $k < -3J$ and in between the frustrated and the trivial paramagnetic phases. 

Our results have implications for several related questions that have been recently studied, and which in turn suggest avenues for future research. These are as follows: 

\smallskip
\noindent
$\bullet$~{\em Classical {TPM} in a field and coupled {TPMs}.} The classical TPMz was first studied in Ref.~\cite{sasa2010thermodynamic}. While our focus here was on ground states and nonthermal properties, for most of the phases at the $k/J<0$ region of the TPMz we confirm the observation that the entropy density of the ``irregularly ordered'' ground states vanishes in the thermodynamic limit. This is true except at special points such as $k=-3J$, cf.\ Fig.~\ref{fig:Degeneracy_scaling_at_PhTr}, or $k=-J$, but also some of the classical frustrated phases studied in Appx.~\ref{appendix:all-models}. 

The phase diagram of the TPMz also describes the behavior of two coupled TPM models \cite{garrahan2014transition,turner2015overlap}, where the order parameter is the overlap between the copies. Both the TPMz \cite{sasa2010thermodynamic} and the coupled TPMs \cite{garrahan2014transition,turner2015overlap} have a first-order transition at low temperatures that, in the limit of zero temperature, approaches $k = 0^+$. However, Refs.~\cite{sasa2010thermodynamic,garrahan2014transition,turner2015overlap} studied systems with PBC and sizes a power of two, which for finite systems have a single classical ground state \cite{sfairopoulos2023boundary}. So a question is how to reconcile this fact with a (classical) ground state singularity at $T=0$. A possible answer is provided by the arguments in Appx.~\ref{appendix:all-models}; in the thermodynamic limit, the minima of the frustrated phase with the largest $M_z$ approach as $L$ increases the $k=0$ point, thus leading to an effective nontrivial ground state degeneracy even for the TPM. As a result, the coexistence of Refs.~\cite{sasa2010thermodynamic,garrahan2014transition,turner2015overlap} would extend to $T=0$ as a consequence of the frustration which is manifest in finite systems at $k<0$.

\smallskip
\noindent
$\bullet$~{\em Dualities in the {TPMz} and {qTPM}.} The observation above should also have implications for the classical duality of the partition sum of the TPMz \cite{sasa2010thermodynamic,garrahan2014transition,turner2015overlap} and the quantum duality of the qTPM \cite{vasiloiu2020trajectory}. Focusing on the qTPM, the current view on quantum dualities \cite{2016_aasen,aasen2020topological,lootens2023dualities,2024_Lootens_2,2024_Lootens,seiberg2024non-invertible,seiberg2024majorana} 
focuses on a non-invertible symmetry operator, $\mathbb{D}$, 
connecting two phases \cite{shao2024whats}. This operator is related to the self-duality by gauging a symmetry of the model. For PBC and a linear size which is a power of two, the TPM has a unique symmetric classical minimum and no nontrivial symmetry apart from the translations. This would indicate the existence of an extra noninvertible symmetry on the self-dual point. Another scenario though, as in the point above, could be that the extra frustrated minima that accumulate at $k = 0^-$ in the large size limit might give rise to a nontrivial symmetry group in the thermodynamic limit that would be related to $\mathbb{D}$. This interplay between dualities and frustrated configurations in plaquette models like the qTPM will be an interesting topic for further study. On similar grounds, it would also be interesting to revisit the classical dualities of the TPMz, as in Refs.~\cite{garrahan2014transition,turner2015overlap}. 

\smallskip
\noindent
$\bullet$~{\em Other plaquette models.} 
It is interesting to compare our results here with other plaquette models. For example, the quantum Baxter-Wu model \cite{baxter2016exactly} in a longitudinal field shows similar effects of frustration and magnetization plateaus for negative $k$. We also expect rich behavior in the ``Rule 150'' plaquette model of Refs.~\cite{devakul2019classifying,you2018subsystem,2020_Tantivasadakarn,stephen2022fractionalization,sfairopoulos2023cellular}. Other systems to consider are the plaquette models related to non-linear CA rules, see Ref.~\cite{sfairopoulos2024in-preparation} for details (for the classical models we expect that their effective defect description gives rise to dualities in their thermal partition sum similar to that of the TPMz \cite{sasa2010thermodynamic,garrahan2014transition,turner2015overlap}). Our approach can also be extended to three-dimensional models, specifically the ``square pyramid model'' \cite{heringa1989phase,turner2015overlap,sfairopoulos2023cellular}, the X-cube model \cite{weinstein2020absence} and the models of Ref.~\cite{canossa2024exotic}. Lastly, note the similarities of the phase diagram of the classical BW model to Fig.~\ref{fig:QuantumPhaseDiagram} \cite{2013_Velonakis}.

\smallskip
\noindent
$\bullet$~{\em Classical stochastic dynamics and large deviations.} The classical TPM is of interest due to its slow glassy relaxation at low temperatures \cite{garrahan2000glassiness,garrahan2002glassiness}. This is understood to be a consequence of the effective kinetic constraints on the motion of defects, which gives rise to dynamical heterogeneity and to large deviation phase transitions \cite{turner2015overlap}. An interesting question is what role the multiplicity of frustrated minima we found here for $k<0$ in the TPMz plays in the glassy dynamics of the TPM. Also, the dynamical phase diagram of KCMs with soft constraints, like the soft FA model \cite{elmatad2010finite-temperature} for positive biasing towards the inactive phase, is reminiscent of the static phase diagram of the TPMz for positive magnetic field \cite{sasa2010thermodynamic,garrahan2014transition,turner2015overlap}. One may wonder if there is also an analogy between the statics of the TPMz for negative field with KCMs with negative activity biasing (noting that for example, in the East model, that region of the dynamical phase diagram also displays what appear to be frustrated states \cite{jack2013large,banuls2019using} with similar patterns as the $3P \times 3Q$ sizes here). As for KCMs the glassiness comes from the dynamical coexistence of an active and an inactive phase, could the glassiness of the dynamics of the plaquette models come from the static coexistence of a frustrated and a trivial phase?

\smallskip
\noindent
$\bullet$~{\em Slow quantum dynamics, fractons, and error correction in the qTPMz.} 
Geometrical frustration has been suggested as a mechanism for disorder-free localization \cite{mcclarty2020disorder-free} and constrained models are known to exhibit slow relaxation and non-thermal eigenstates \cite{horssen2015dynamics,lan2018quantum,
turner2018weak,pancotti2020quantum,roy2020strong,valencia-tortora2022kinetically, 
bertini2024localized,stephen2024ergodicity,causer2024quantum}. We expect the qTPMz, at least near both $k/J=0$ and $k/J=-3$, to display some of this behavior in its dynamics. Furthermore, based on the form of Eq.~\ref{QTPMzv2}, it would be intriguing to check whether the results of Ref.~\cite{stephen2024ergodicity} apply to the qTPMz too.
Quantum plaquette models are also relevant to the study of fractons 
\cite{chamon2005quantum,castelnovo2012topological,devakul2019fractal,pretko2020fracton,seiberg2020exotic,seiberg2021exotic} and recently have also found applications in error correction, see for example Refs.~\cite{rakovszky2023physicsgoodldpccodes,rakovszky2024physicsgoodldpccodes,hong2024quantummemorynonzerotemperature}.

\smallskip
\noindent
$\bullet$~{\em Improvements on our results.} While we are confident on the ground state phase structure of the qTPMz we described here, improving on our numerics could help solidify our observations. While we focused on the exponential degeneracy of the classical minima at the $k=-3J$ point due to its effect on the qTPMz around that point, there are other isolated locations of the classical TPMz with exponential number of ground states for some sequences of system sizes. We have seen that one such point is the $k=-J$ point, according to our numerics for sizes $L \neq 3P$ (not shown). While the quantum fluctuations around this point appear small for most system size sequences, for $L=4$ they do appear to be strong, destroying the classical frustrated phase order for all $h/J \neq 0$. It would thus be interesting to study the behavior of the qTPMz around it more exhaustively. Lastly, further intuition and results would be needed for the clarification of the existence or not of a spin liquid regime.

\smallskip 
We hope to report on some of the above questions in future works. 

\begin{acknowledgments}
       We thank L. Causer, J. Mair, S. Powell for collaborations on related projects and I. Lesanovsky, J. C\^{o}t\'{e} and J. Biamonte for helpful discussions. We want to particularly thank A. Gammon-Smith and C. Castelnovo for valuable discussions and feedback on the manuscript.
       We acknowledge financial support from EPSRC Grant No.\ EP/V031201/1.
       Simulations were performed using the University of Nottingham Augusta and Ada HPC clusters (funded by EPSRC Grant EP/T022108/1 and the HPC Midlands+ consortium). 
\end{acknowledgments}

\bibliography{bibliography-06092024}

\appendix
\section{Proof of $M_z=-1/3$}{\label{appendix:proof}}

In this Appendix we study the magnetization plateaus of the classical TPMz for system sizes $N = 3P \times 3Q$ (with PBC) and for $k\in [-3J, 0]$.

The minima of \er{TPMz} for $k\in [-3J, 0]$ are obtained from configurations that minimize collectively the plaquettes of the TPM and have the lowest (i.e., most negative) magnetization. For the $N = 3P \times 3Q$ system sizes, all plaquette terms are simultaneously minimized and so the frustration effects are also trivialized. Obtaining these becomes a tiling problem. Consider dividing the $N = 3P \times 3Q$ lattice into $P \times Q$ contiguous blocks of size $3 \times 3$. In isolation, each of these blocks is a TPMz with open boundary conditions (OBC), where the interaction energy is only in the four plaquettes fully contained in the block. Such $3 \times 3$ OBC TPMz has $2^5$ possible ground state configurations with no excited plaquettes for its ground states: one of these has magnetization density $1$, one magnetization density $7/9$, one $5/9$, three $1/3$, seven  $1/9$, eleven $-1/9$, seven $-1/3$, and one $-5/9$. We obtain a ground state of the $N = 3P \times 3Q$ system by tiling the space with these $3 \times 3$ configurations, where the matching rule is that the plaquettes straddling the junctures of the tiles also need to be unexcited. This severely restricts which tiles are relevant for the tiling. Specifically, the tile with magnetization $-5/9$ cannot be used to cover the space on its own without giving rise to excited plaquettes, and to make it ``tilable'' it has to be ``screened'' by 2 tiles with magnetization higher than $-1/3$, thus giving rise to a total magnetization larger than $-1/3$. In contrast, three of the seven tiles with magnetization $-1/3$, specifically the arrangement of Fig.~\ref{fig:tiles}(k) and its two translation invariant partners, can be used to periodically cover the space, giving rise to the threefold-degenerate minima of the TPMz for $k \rightarrow 0^-$, cf.\ Fig.~\ref{fig:plateaus6and9}. 

The approach above can be similarly applied to the $k=-3J$ point. In this limit, similar arguments show that the ground states with the highest magnetization are the translationally symmetric ones of Fig.~\ref{fig:tiles}k, too. The argument concerns finding the tiles that lead to the fully packed configurations of trimers for the $3\times 3$ blocks for $k \rightarrow -3J^+$, cf. Sec.~\ref{k=-3J}.  

Then, starting from the $k=0$ point, the ground states for an arbitrarily small $k$ with $k<0$ will be those states with the largest allowed slope for their classical energy functions, $E/N = -1 - k M_z$. Now we perturb the model with $E = \lim_{k\rightarrow - 3J^{+}}E$ and see that the only ground states are the same ones as for the $E = \lim_{k\rightarrow 0^{-}}E$ case. Since all energy states have energies which are first-order polynomials with respect to the longitudinal field, we conclude that these states will necessarily be the only ground states of the model for the $k\in (-3J, 0)$ section. Similar procedures can be followed for the identification of the highest and lowest magnetization plateaus for other $N = kP \times kQ$ system sizes, with $k \geq 3$, although in these cases the two of them are different from each other. This concludes the proof.
\begin{figure*}
    \centering
    \begin{subfigure}[b]{0.24\textwidth}
        \includegraphics[width=\linewidth]{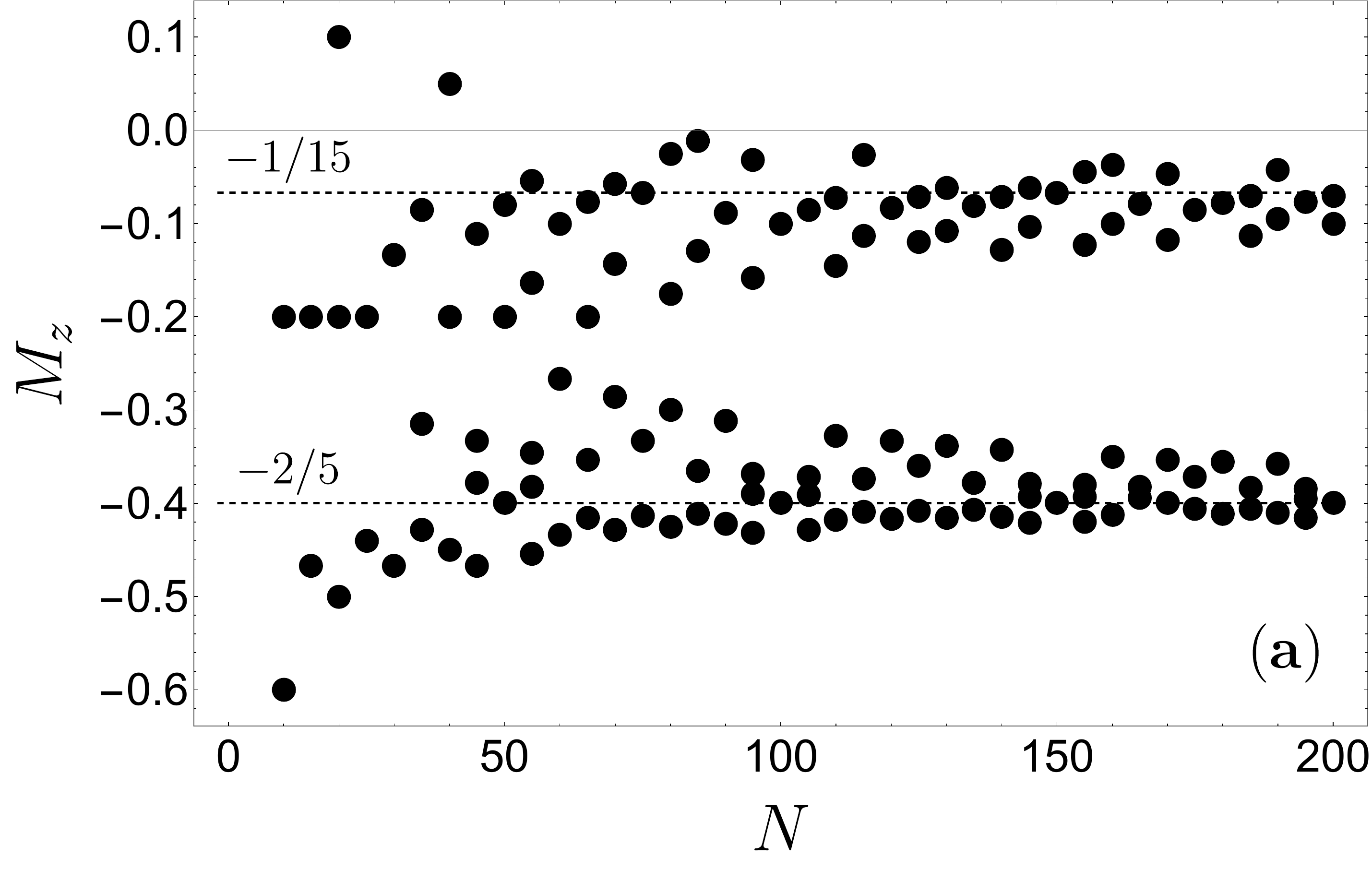}
     \end{subfigure}
     \begin{subfigure}[b]{0.24\textwidth} 
        \includegraphics[width=\linewidth]{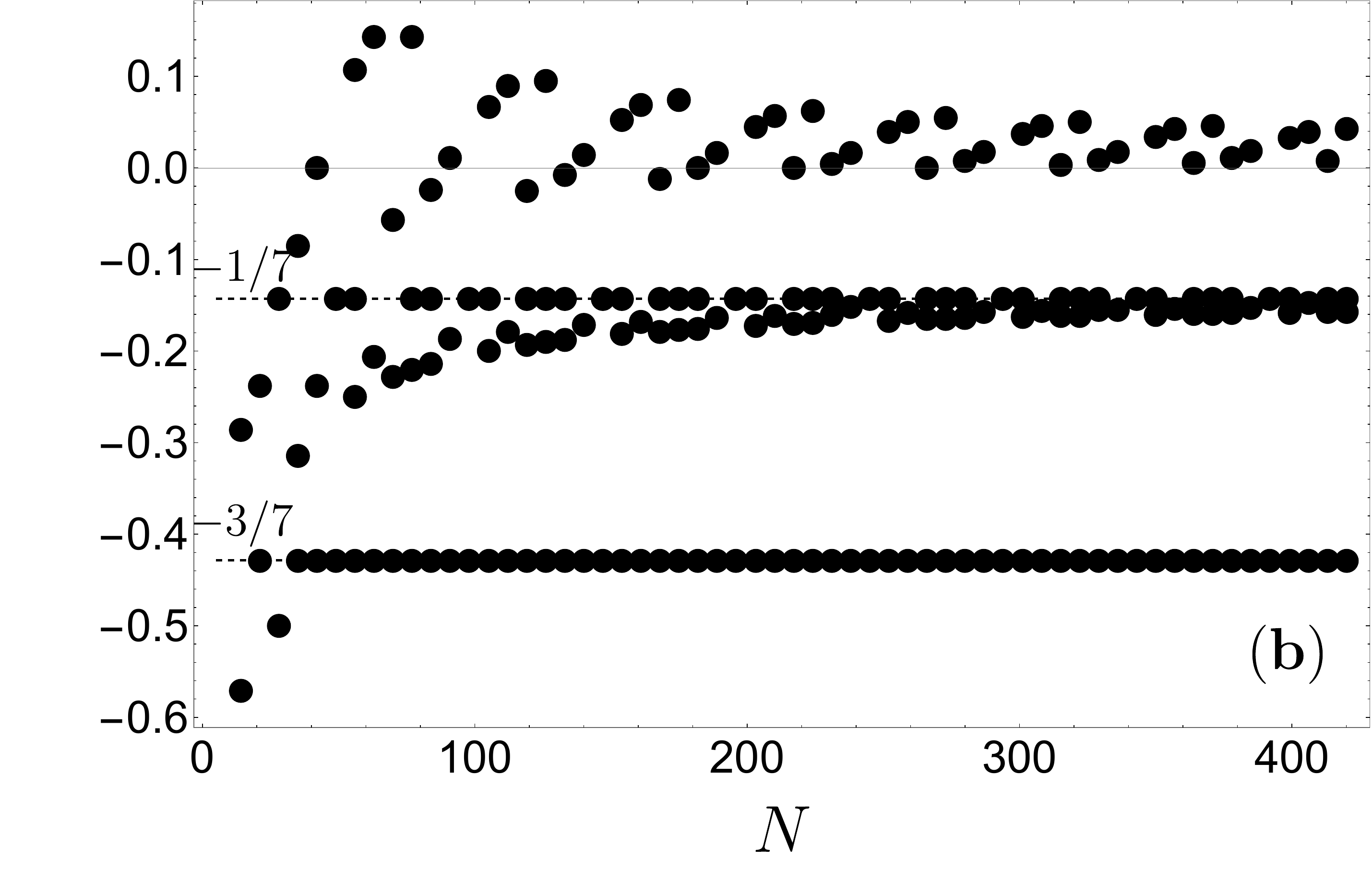}
     \end{subfigure}
     \begin{subfigure}[b]{0.24\textwidth} 
      \includegraphics[width=\linewidth]{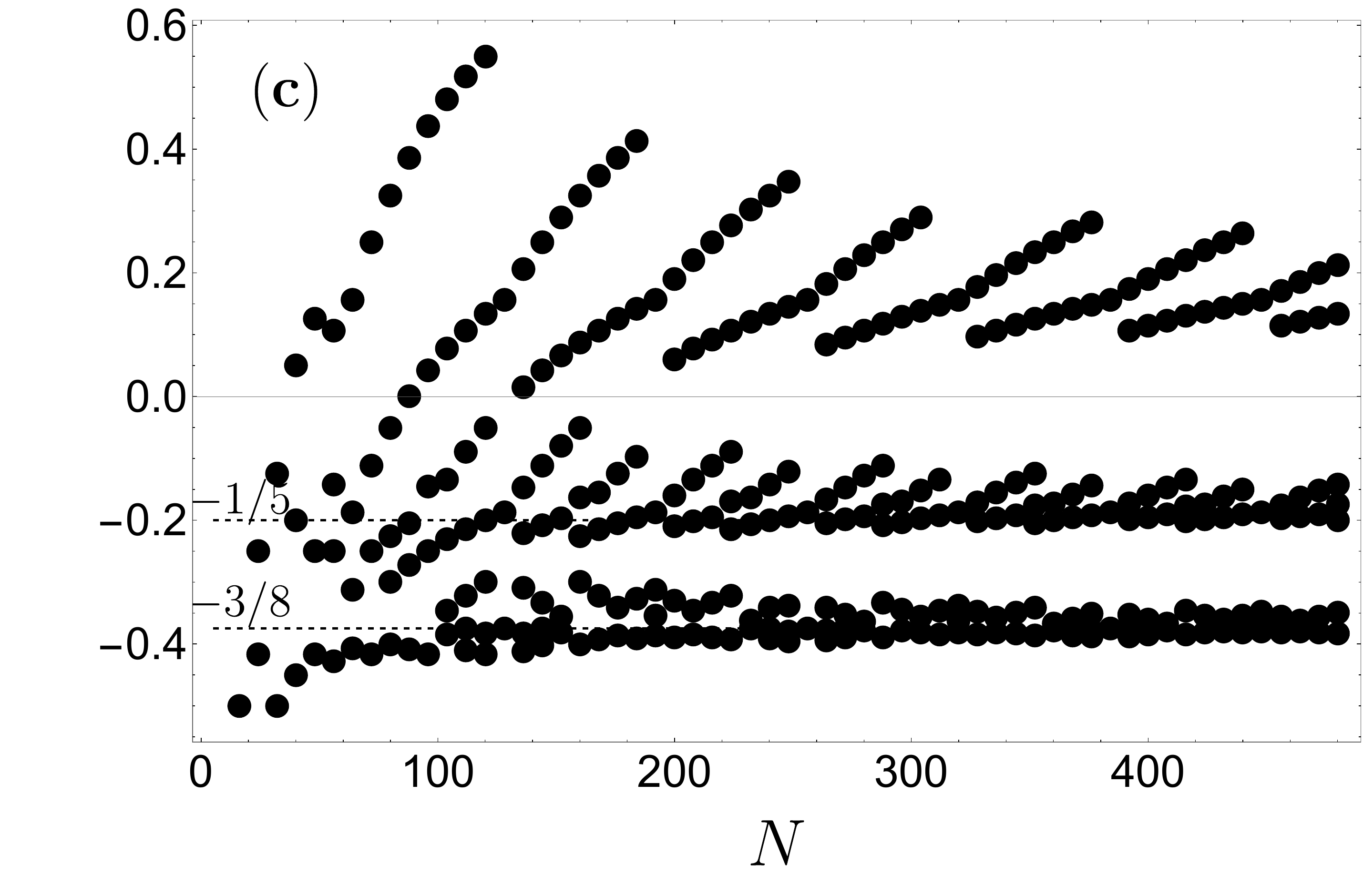}
     \end{subfigure}
     \begin{subfigure}[b]{0.24\textwidth} 
        \includegraphics[width=\linewidth]{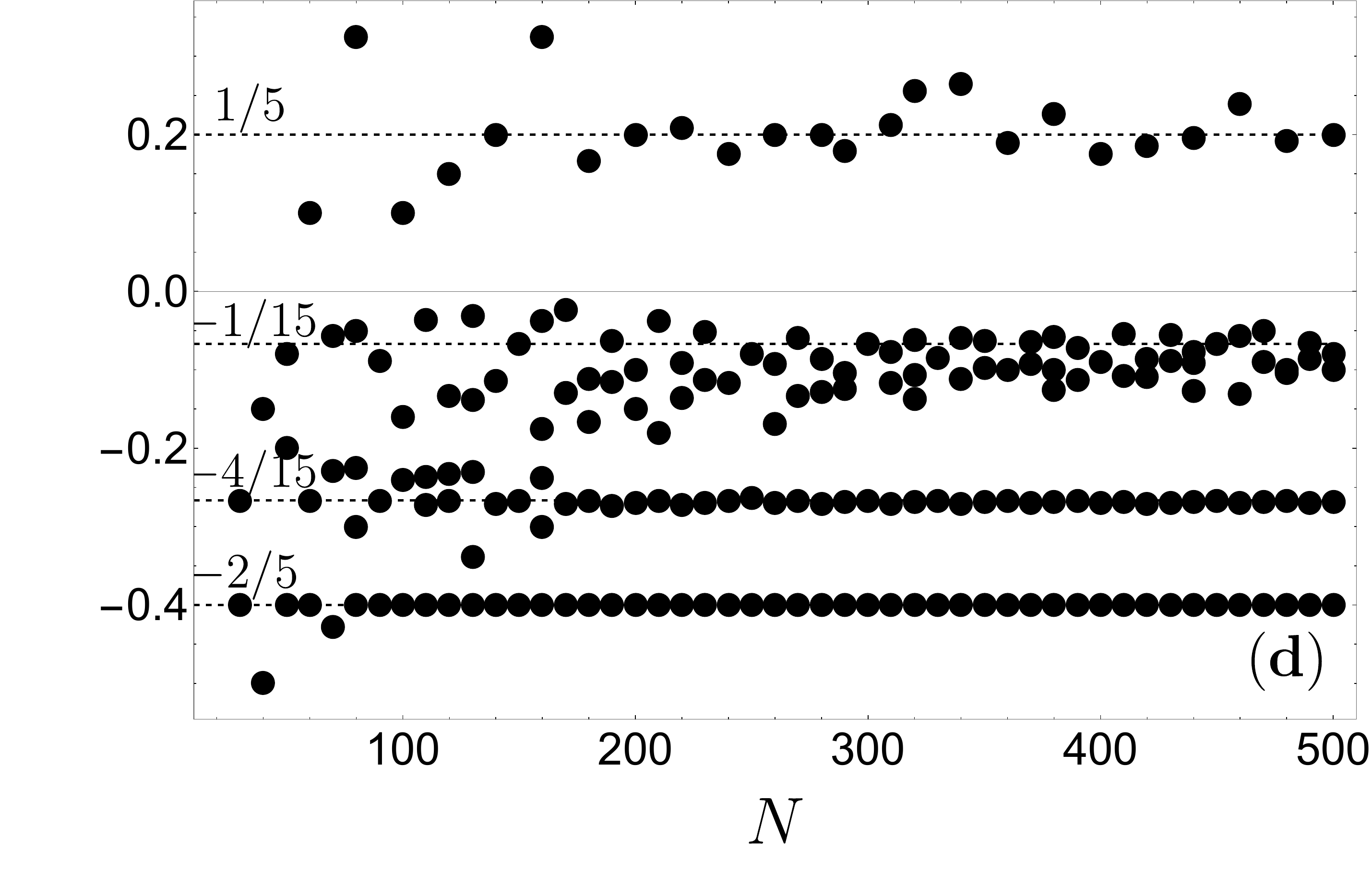}
     \end{subfigure}
     \caption{(a) Magnetization plateaus for $N=L \times M$ with $L = 5$ fixed. (b) Same for $L=7$. (c) Same for $L = 8$. (d) Same for $L = 10$.}
     \label{fig:plateaus-5-7-8-10}
\end{figure*}

\begin{figure*}[t]
    \centering
    \begin{subfigure}{0.3\textwidth} 
      \includegraphics[width=\linewidth]{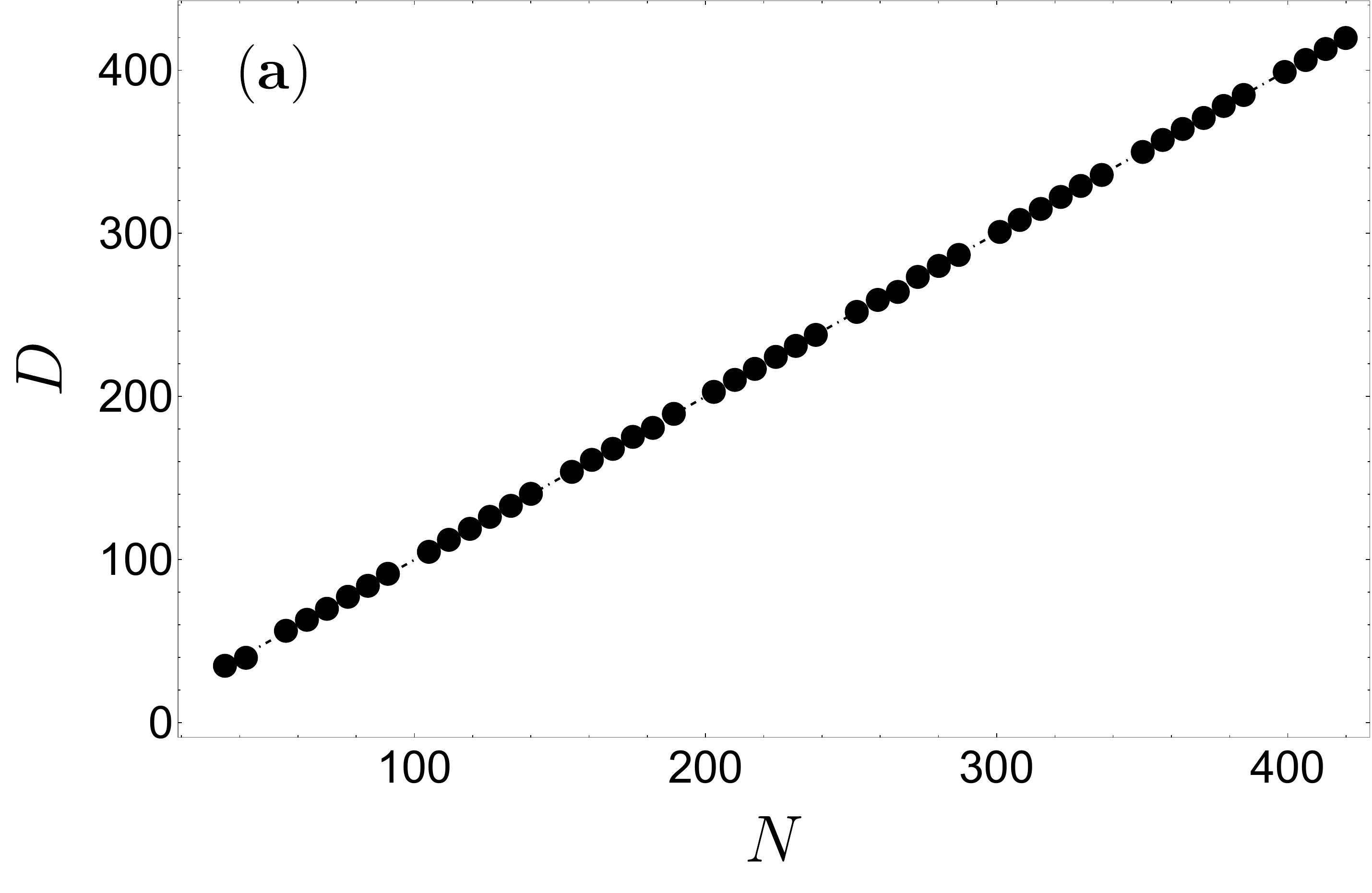}
    \end{subfigure}
    \begin{subfigure}{0.3\textwidth}
        \includegraphics[width=\linewidth]{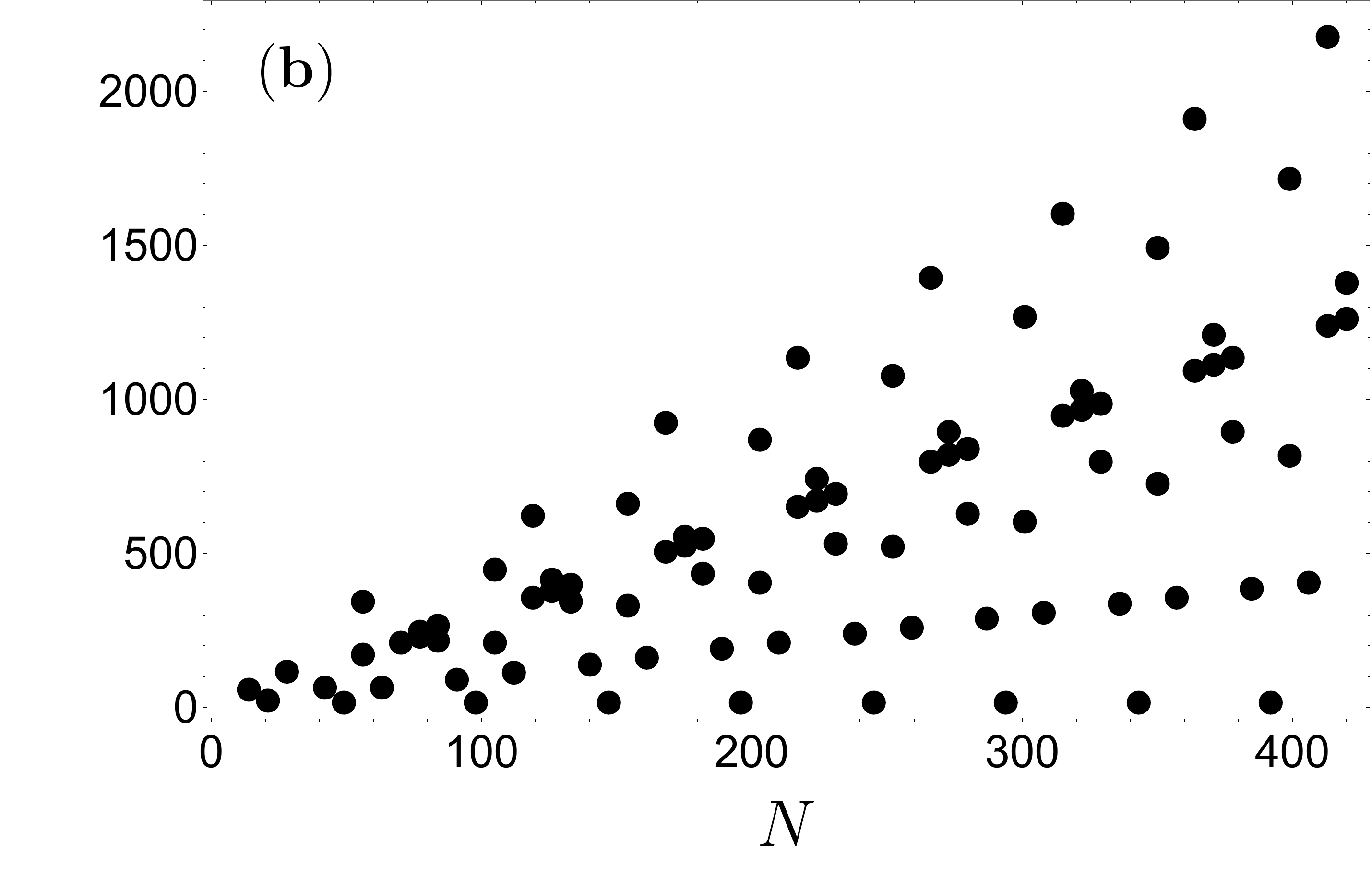}
     \end{subfigure}
     \begin{subfigure}{0.3\textwidth} 
        \includegraphics[width=\linewidth]{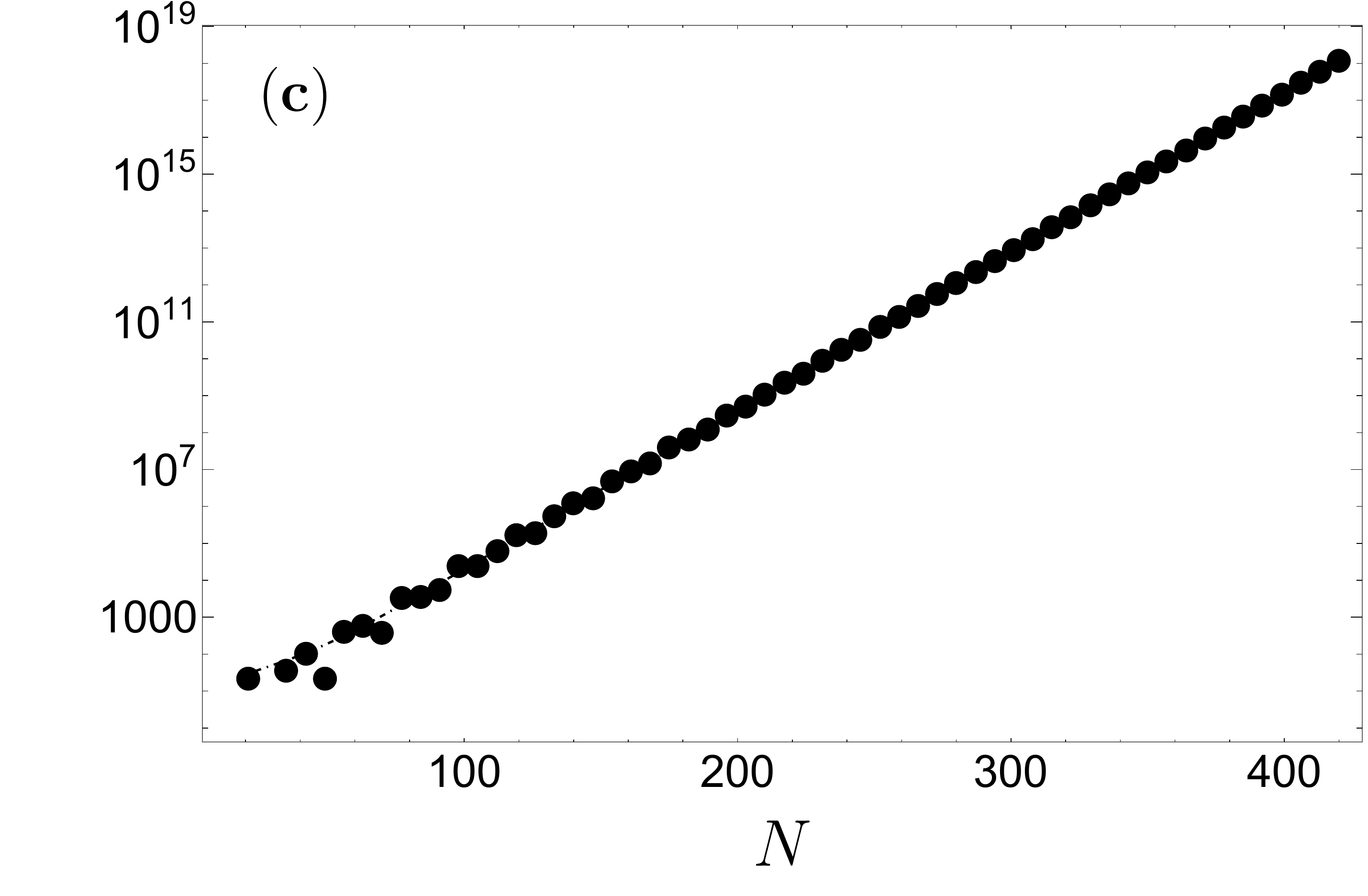}
     \end{subfigure}
     \begin{subfigure}{0.3\textwidth} 
      \includegraphics[width=\linewidth]{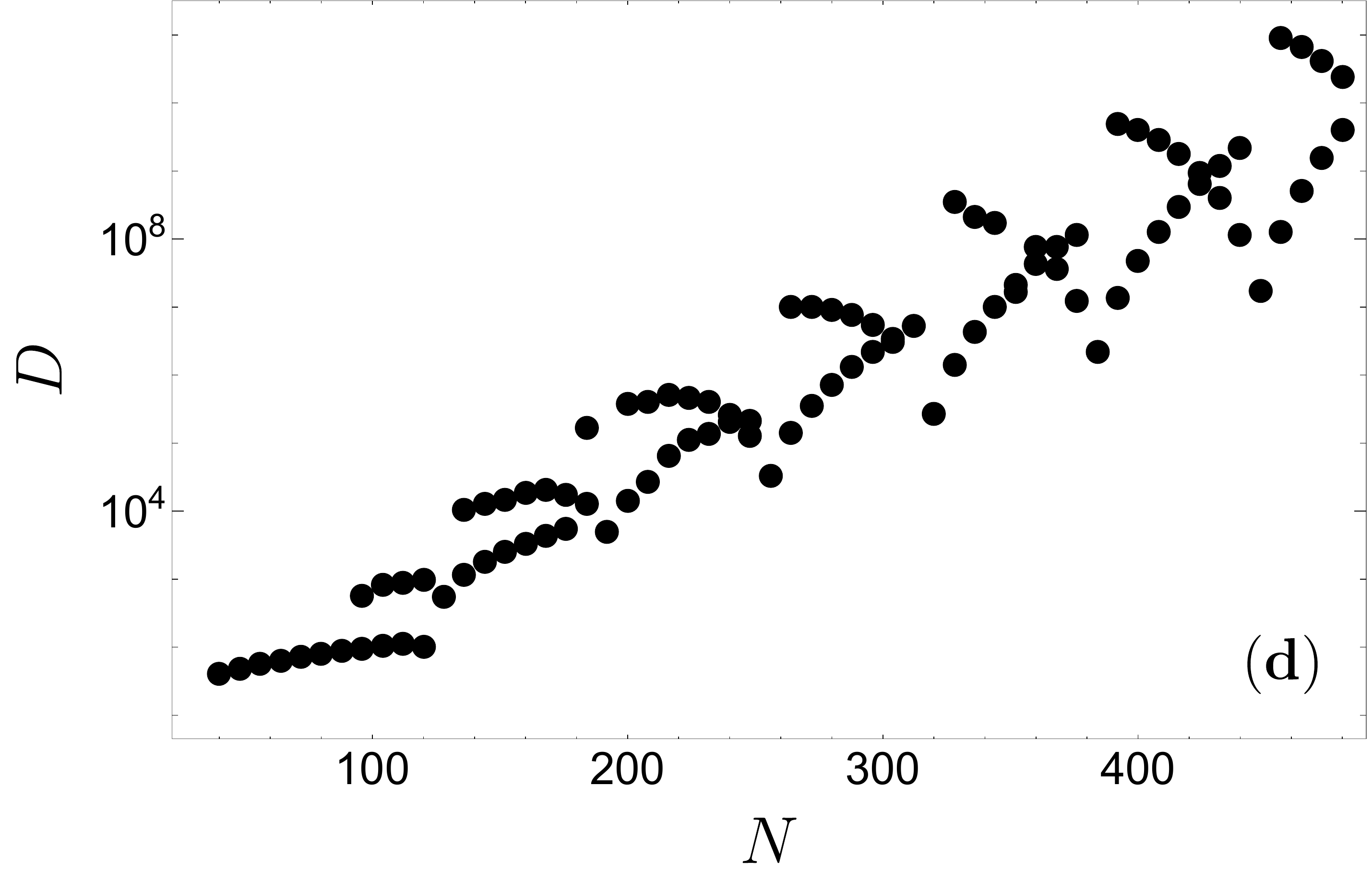}
      \end{subfigure}
      \begin{subfigure}{0.3\textwidth} 
         \includegraphics[width=\linewidth]{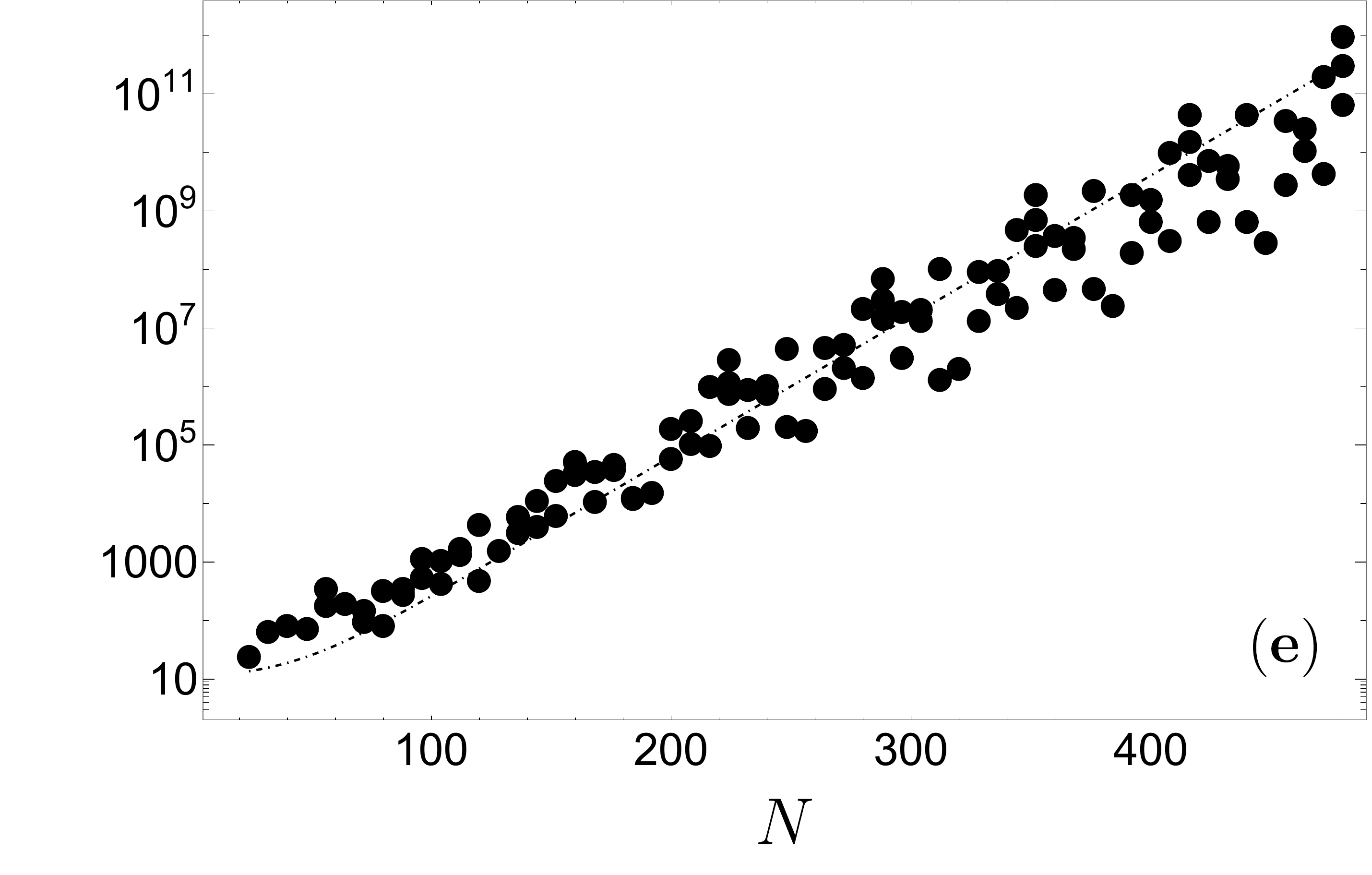}
      \end{subfigure}
      \begin{subfigure}{0.3\textwidth} 
         \includegraphics[width=\linewidth]{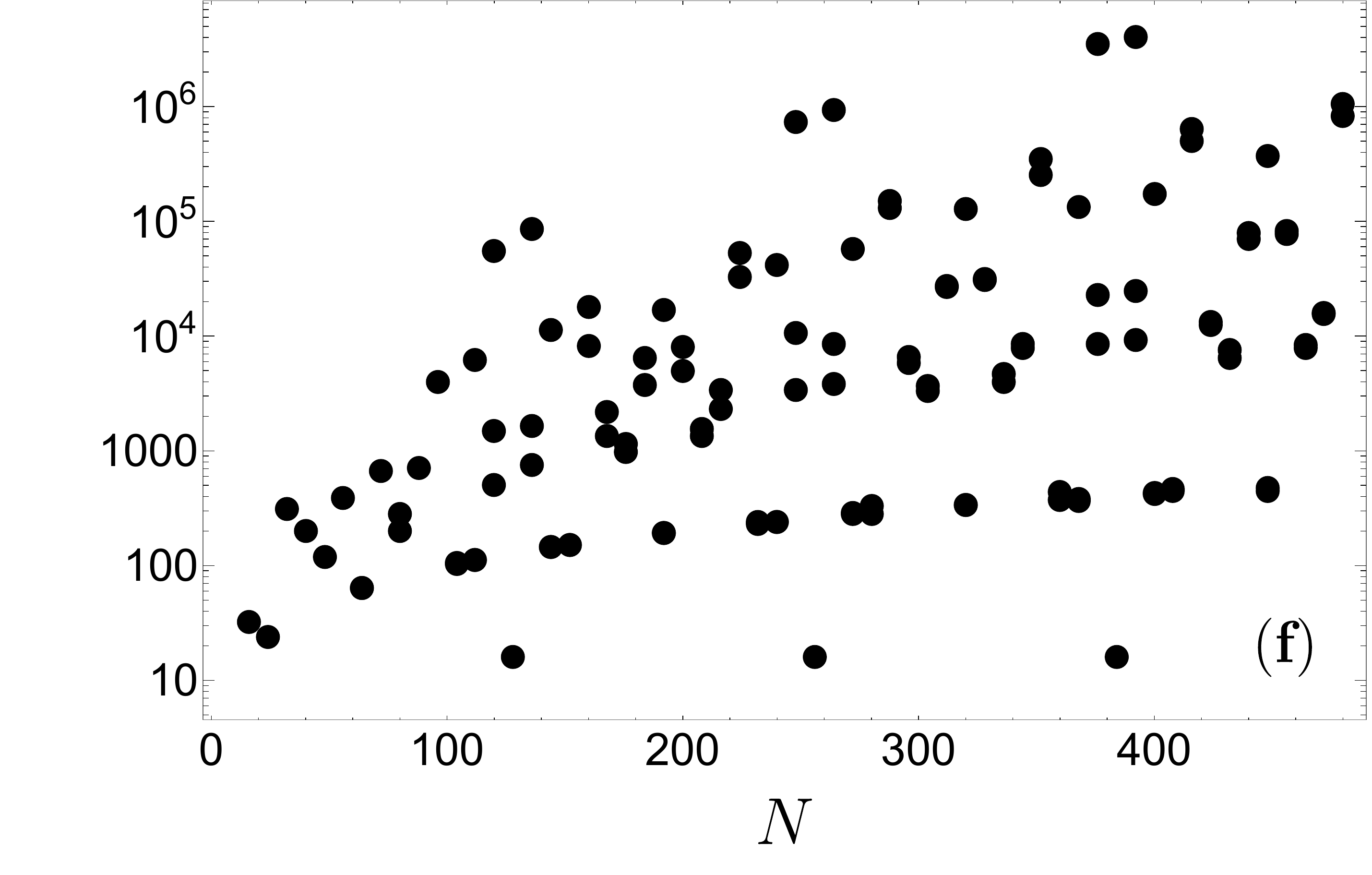}
      \end{subfigure}
      \begin{subfigure}{0.24\textwidth} 
         \includegraphics[width=\linewidth]{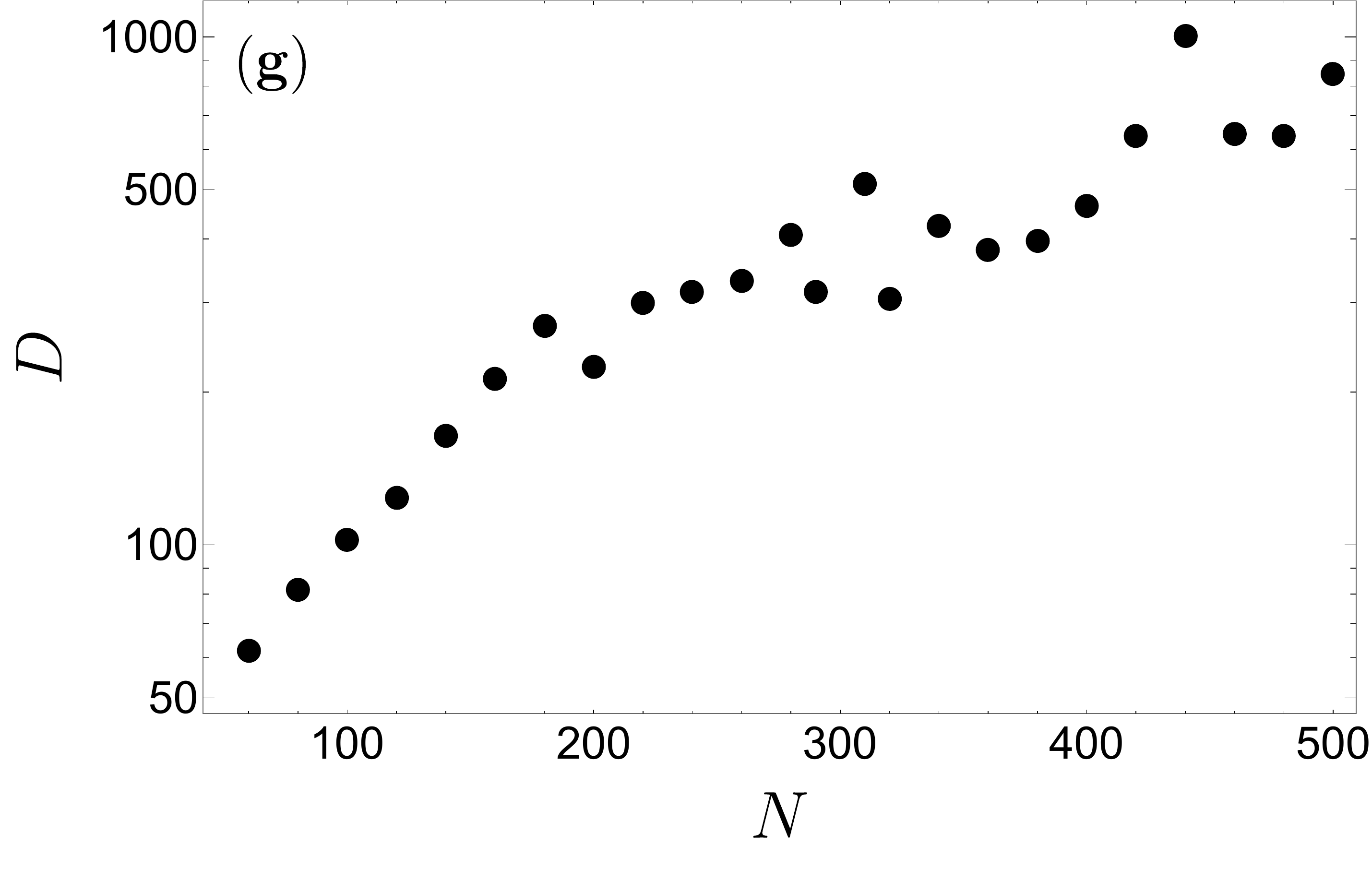}
      \end{subfigure}
      \begin{subfigure}{0.24\textwidth} 
         \includegraphics[width=\linewidth]{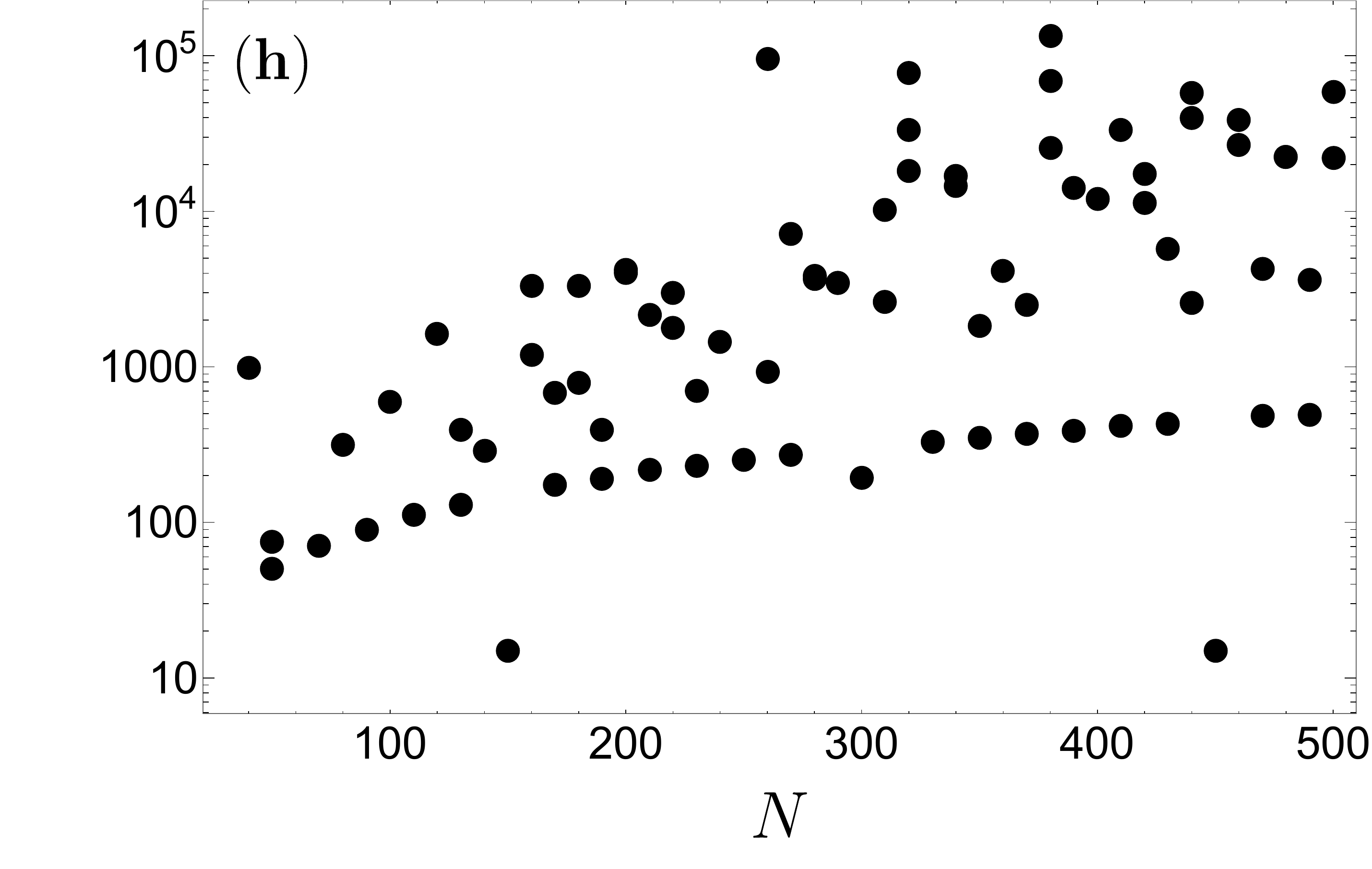}
      \end{subfigure}
      \begin{subfigure}{0.24\textwidth} 
         \includegraphics[width=\linewidth]{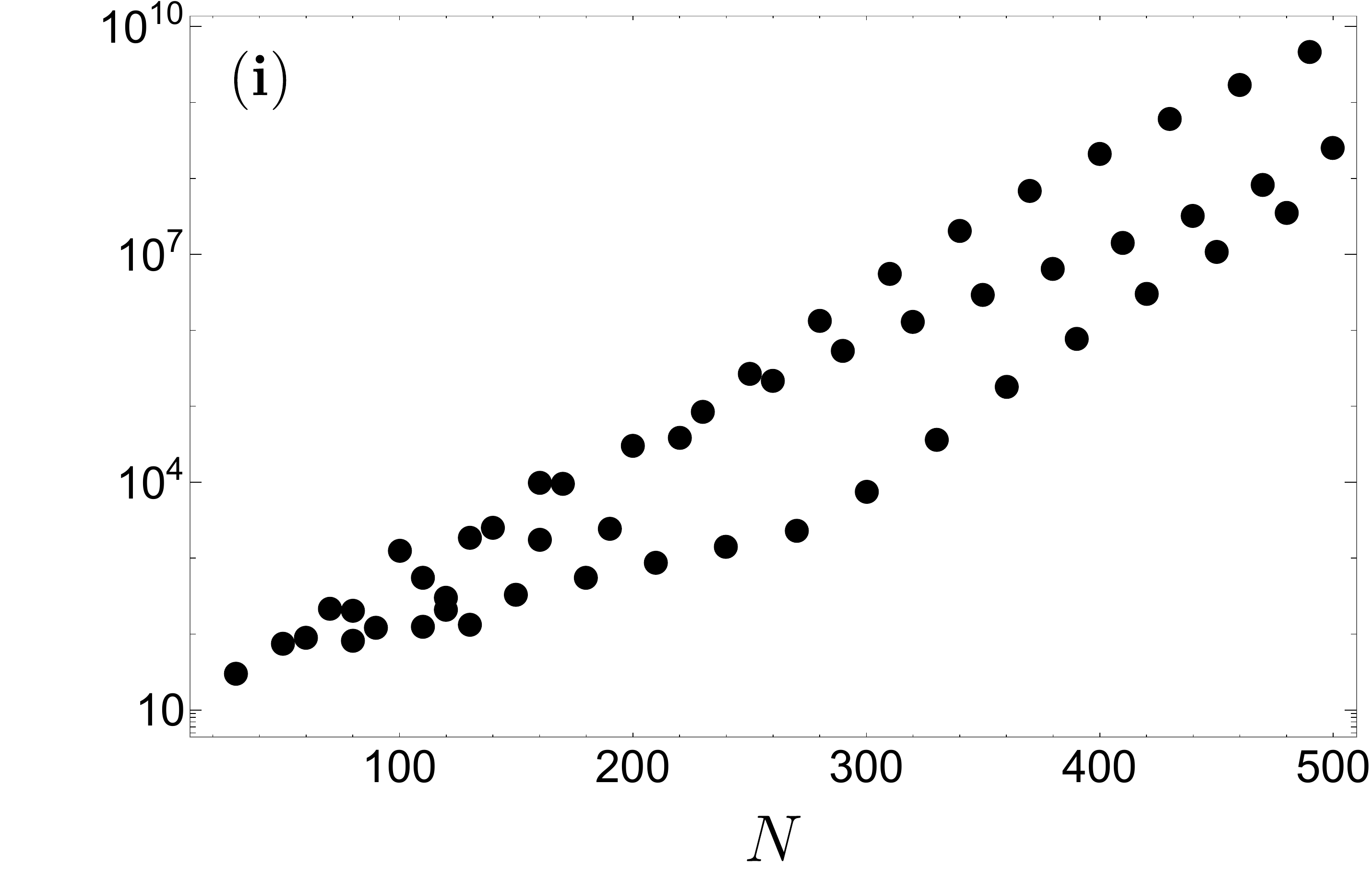}
      \end{subfigure}
      \begin{subfigure}{0.24\textwidth} 
         \includegraphics[width=\linewidth]{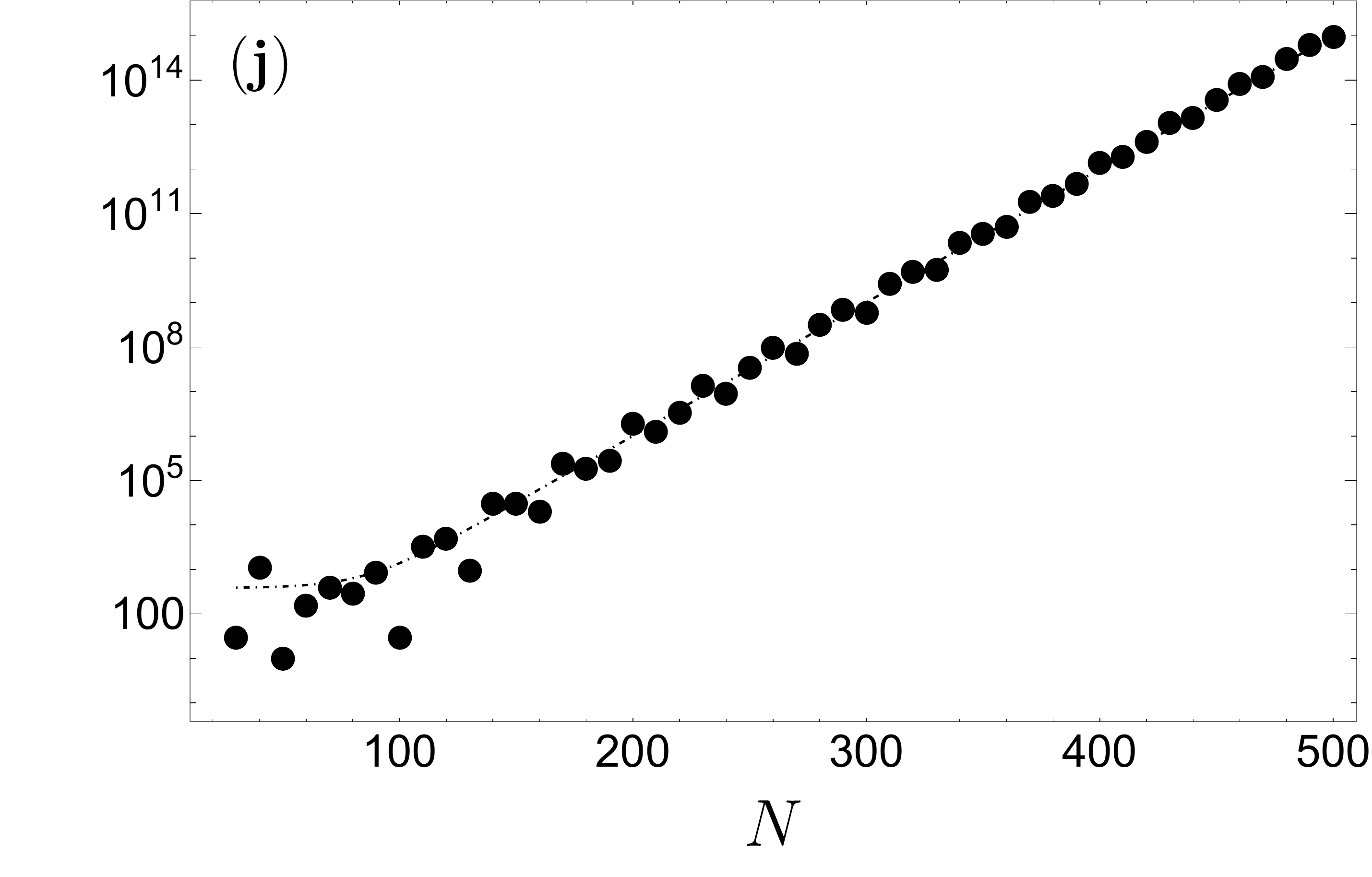}
      \end{subfigure}
     \caption{(a) Plateau degeneracies for $L = 7$ fixed and for the plateaus with $M_z \gtrsim -0.11$. (b) Same for $M_z \approx -1/7$ (linear fit). (c) Same for $M_z \approx -3/7$ (exponential fit).(d) Plateau degeneracies for $L = 8$ and $M_z > 0.0$. (e) Same for $M_z \approx -1/5$ (exponential fit). (f) Same for $M_z \approx -3/8$. (g) Plateau degeneracies for $L = 10$ and $M_z > 0.0$. (h) Same for $M_z \approx -1/15$. (i) Same for $M_z \approx -4/15$. (j) Same for $M_z \approx -2/5$ (exponential fit).}
  \label{fig:degeneracies-7-8-10}
\end{figure*}

\section{Further results for the magnetization plateaus}{\label{appendix:all-models}}

 In the main text we discussed the magnetization plateaus of the minima of the TPMz in the regime $[-3J, 0]$ for systems of size $N = L \times M$ for two values of $L$, $L=6$ and $L=9$. Here we present similar results for $L \in \{5, 7, 8, 10\}$.
 In analogy to Figs.~\ref{fig:plateaus6and9}(a,b), we show the values of the plateaus as a function of system size for $L=\{5,7, 8, 10\}$ in Fig.~\ref{fig:plateaus-5-7-8-10}. 

 For $L=5$ and for increasing system size, there are two magnetization plateaus, for $M_{z} = -1/15$ and for $M_{z}=-2/5$ around which intermediate magnetization plateaus seem to coalesce. The width of both plateaus shows stability to increasing the system size, verifying their presence towards the thermodynamic limit. At the $k=0$ point, in the thermodynamic limit, the FM and the frustrated phase ground states coexist. A similar situation can be verified from the exact ground states for the classical TPM for a given system size $N=5 \times 15$ (which is the smallest finite system size with a nontrivial ground state degeneracy); one ground state has magnetization $M_z=1$ (the trivial FM state) and all other states have $M_z = -1/15$. The difference between the cases $M=15Q$ and $M\neq15Q$, for $Q \in \mathbb{Z}$, for $L=5$ lies in the degeneracy of their respective ground states.

 For $L=7$ the convergence to the thermodynamic limit is slower. However, the conclusions above apply as well. The plateaus with $M_z \gtrsim 0$ approach a value very close to $M_z = 0.02$ but their $k$-range decreases with increasing system size, suggesting that these plateaus disappear in the large size limit. The only stable plateaus are those around $M_z = -1/7$ and $M_z = -3/7$. In the large size limit, for $k=0$ we observe the coexistence of the ground states from the $M_z = -1/7$ and $M_z=1$ phases, in addition to the ones coming from the $M_z \approx 0.0$ plateau. We verify this statement from the ground states of the $N=7\times7$ system, which plays the same role as for the $L=5$ case. We find ground states with magnetization $M_z = 1$ (FM state), $M_z=-1/7$ (frustrated states) and $M_z = 1/49$ (extra degenerate states). The differences between $M=7Q$ and $M\neq7Q$, for $Q \in \mathbb{Z}$, for $L=7$ lie in the degeneracy of their respective ground states, as shown in Fig.~\ref{fig:degeneracies-7-8-10}(a-c). 

 For $L=10$ we find that the width of the magnetization plateaus for $M_z \approx 1/5$ decreases with system size.  
 Approaching the thermodynamic limit at $k \rightarrow 0$, we can distinguish the two phases, the FM one for positive $k$ and the frustrated phase with $M_z = -1/15$ for negative $k$. As above, the ground state degeneracy of the classical TPM can be confirmed in the case of the system size $N=10\times 15$; there are ground states with magnetization $M_z = 1$, $M_z = 1/5$ and $M_z = -1/15$. 
 An interesting feature of the model is observed for the $M_z=-2/5$ plateau. The existence of this plateau for both $L=5$ and $L=10$ is not accidental, but comes from the tiling of the lattices with $L=5Q$ near $k=-3J$. The degeneracies of these plateaus are shown in Fig.~\ref{fig:degeneracies-7-8-10}(g-j).

 A different behavior is observed for $L=8$. The magnetization plateaus for $M_z>0$ show a similar behavior to the positive plateaus for $L=7$ but with more extended structures. Their magnetization value in the large size limit approaches the value $M_z = 1/8$. Similarly involved magnetization plateau structures are found around the $M_z = -1/5$ and $M_z = -3/8$ plateaus for decreasing $k$. Special care is needed though for $L=8$ and any $L = 2^p$ size for $p \in \mathbb{Z}$. The plateau with the smaller absolute magnetization for a given $L$ never shifts with increasing $M$ and never approaches the $k=0$ point. This argument confirms the ground state degeneracy of the TPM for system sizes with a given $L$ that is a power of two (single ground state). At the same time though, the negative $k$ value where we observe the phase transition to the frustrated phases shifts towards $k=0$ for increasing $L$ (but still a power of two). As a result, we believe that even for linear system sizes a power of two, the $k=0$ point will effectively include effects from the frustrated phases in the thermodynamic limit. We note that in going to large $N$, the limits $L \rightarrow \infty$ and $M \rightarrow \infty$ do not commute. The degeneracies of the plateaus for $L=8$ are given in Fig.~\ref{fig:degeneracies-7-8-10}(d-f).

 Similar arguments hold for the cases of $L=6$ and $L=9$, studied in Sec.~\ref{classical_frustration}. Focusing on the more complicated case of $L=9$, the nontrivial periods of the CA rule 60 give periods of length 3 and 63. The ground states of the TPM for a $9 \times 63Q$ system size have magnetization $11/63$, $-1/63$, $-5/63$ and $-1/3$. Following our previous analysis we would expect the occurrence of plateau ``phases'' around these magnetization values for increasing system sizes. This is indeed the case, as shown in Fig.\ref{fig:plateaus6and9}, although the convergence to those exact values is not as good as for the previos cases studied. All these plateaus, however, are visible with the exemption of the $-1/63$ one. This phenomenon is due to the small system sizes studied compared to the period length of the target magnetization plateaus and we thus expect them to be visible (and converged) for larger systems. The case $L=6$ possesses no particular surprises and the explanation of the appearing plateaus follows exactly from the preceding discussions.

 Fig.~\ref{fig:degeneracies-7-8-10} shows the degeneracy of the minima at the plateaus studied above as a function of $N$ for fixed $L \in \{7, 8, 10\}$ for various ranges of $M_z$. For the $L=7$ and $L=10$ cases for the higher magnetization plateaus we observe a linear-in-size degeneracy. These states will contribute to the $k=0$ point in the thermodynamic limit and give a higher degeneracy for the TPM (while still having a vanishing ground state entropy density, making them unimportant at finite temperatures \cite{newman1999glassy,garrahan2000glassiness}). 
 For the case $L=8$ a distinctly different pattern is observed. Since the higher magnetization plateaus never shift towards the $k=0$ point for a given $L$, they can accomodate any ground state degeneracy scaling, even an exponential one. 

 Some of the (stable thermodynamically) frustrated phases exhibit an exponential entropy scaling; for example, for $L=7$ we find an entropy, $S$,  $S/N \approx 0.1$ for $k=-2.2$ and for a sequence of sizes inside the $M_z = -3/7$ phase, for $L=8$ and the phase with $M_z = -1/5$ the entropy saturates around $S/N \approx 0.04$ and for $L=10$ and for the $M_z = -2/5$ phase $S/N \approx 0.07$.

 Here, we want to summarize the results of this section but also bring them in perspective compared to the results of Sec.~\ref{classical_frustration}. For all linear system sizes studied (with the exemption of systems of size $3P \times 3Q$ and for $L=2^P$ where $P, Q$ integers) we have found a qualitative criterion for when the magnetization plateaus for the largest magnetization(s) shift and approach the $k=0$ point with increasing the vertical system size dimension. This would in effect give a higher ground state degeneracy in the thermodynamic limit for the line with $k=0$ where the classical TPM is located. This criterion can be obtained by: (i) inspecting the periodic orbits of the Rule 60 CA that gives rise to the ground states of the classical TPM \cite{sfairopoulos2023boundary}, (ii) calculating the classical magnetization for those states, (iii) if plateaus with magnetization larger than the minimum magnetization for the states from step (ii) are found, then those plataeus would necessarily have to shrink in the thermodynamic limit and collide to the $k=0$ point, thus giving rise to an increased accidental ground state degeneracy for that point. For the plateau phase with the lowest magnetization, we argue that the specfic value of their magnetization comes from a tiling problem fully constrained by the $k=-3J$ point. To get the value of these plateaus, we would have to construct all the ground states for that point and select the ones with the largest magnetization. All our arguments above follow from the fact that the phase transitions between these phases (assuming the thermodynamic limit) are of first order, as our numerical results indicate and the principle of coexistence at the phase transition points.

 One important and surprising result that we found concerns the difference of the limits $\lim_{L\rightarrow \infty} \lim_{M\rightarrow \infty}$ vs. $\lim_{M\rightarrow \infty} \lim_{L\rightarrow \infty}$ for the case of system sizes a power of two. This is a reminder that care is needed when taking the thermodynamic limit.

 Lastly, we want to comment on the degeneracies of those plateaus in Figs.~\ref{fig:plateaus6and9}-\ref{fig:degeneracies-7-8-10}. Although most cases give rise to plateau degeneracies that are subextensive, there are cases where the ground state degeneracies become extensive, thus leading to a finite thermodynamic entropy. Further study of those situations, the origin of the extensivity of the ground state manifold, its effects on observables of the system and the nature of the phase transitions would be needed.
 
\begin{figure}[t]
   \centering
   \begin{subfigure}[b]{0.32\columnwidth}
      \includegraphics[width=22mm, height=22mm]{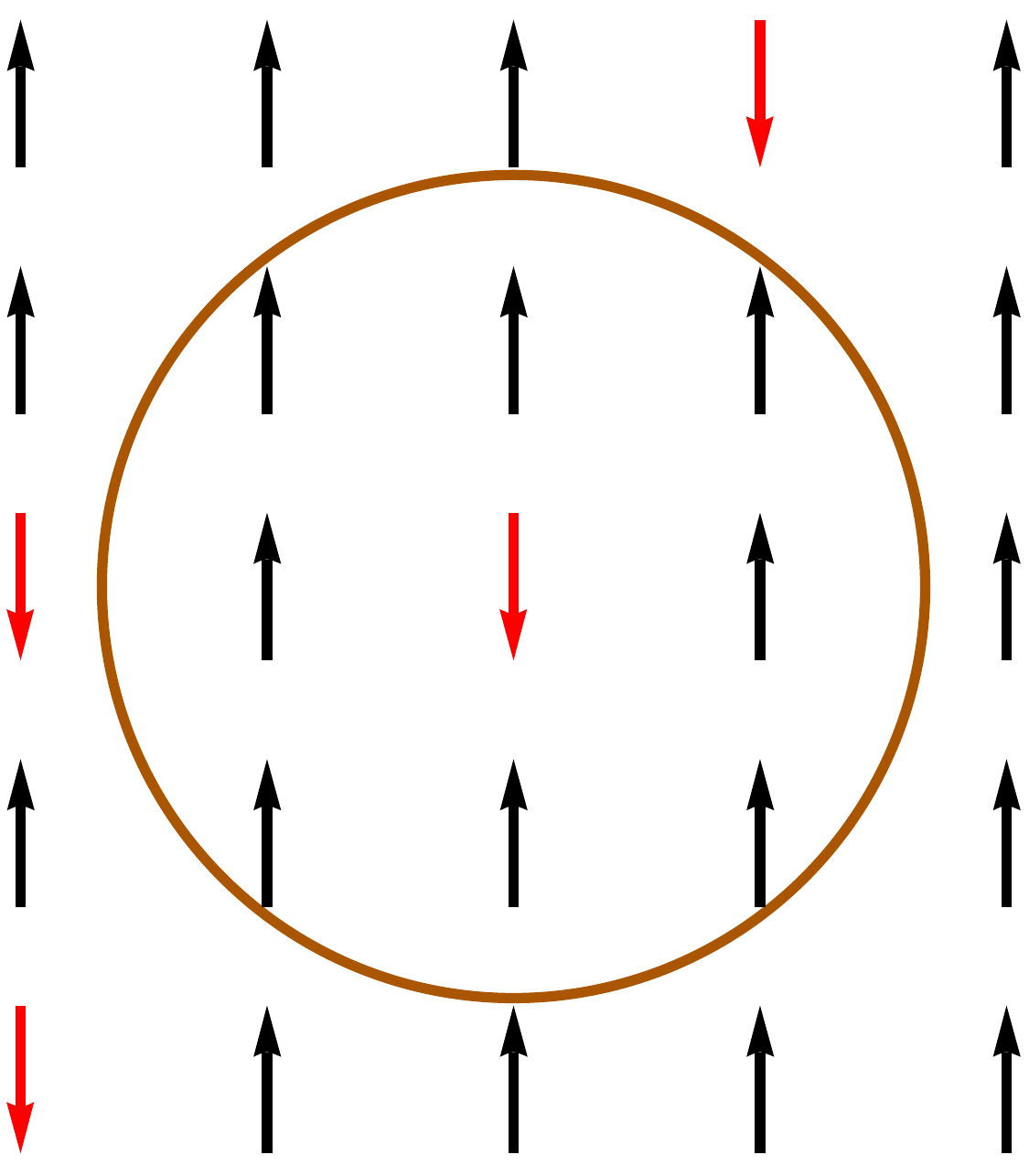}
      \subcaption{}
    \end{subfigure}
    \begin{subfigure}[b]{0.32\columnwidth}
       \includegraphics[width=22mm, height=22mm]{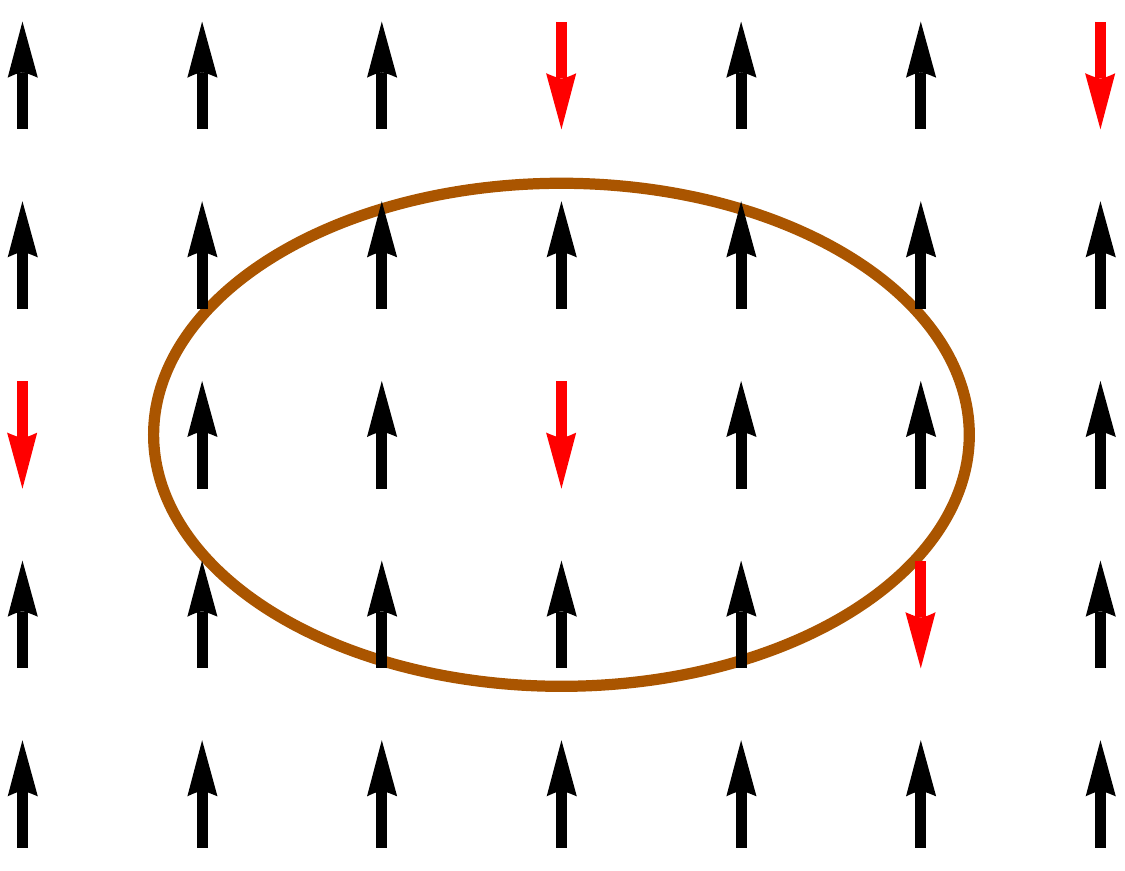}
       \subcaption{}
    \end{subfigure}
    \begin{subfigure}[b]{0.32\columnwidth}
       \includegraphics[width=22mm, height=22mm]{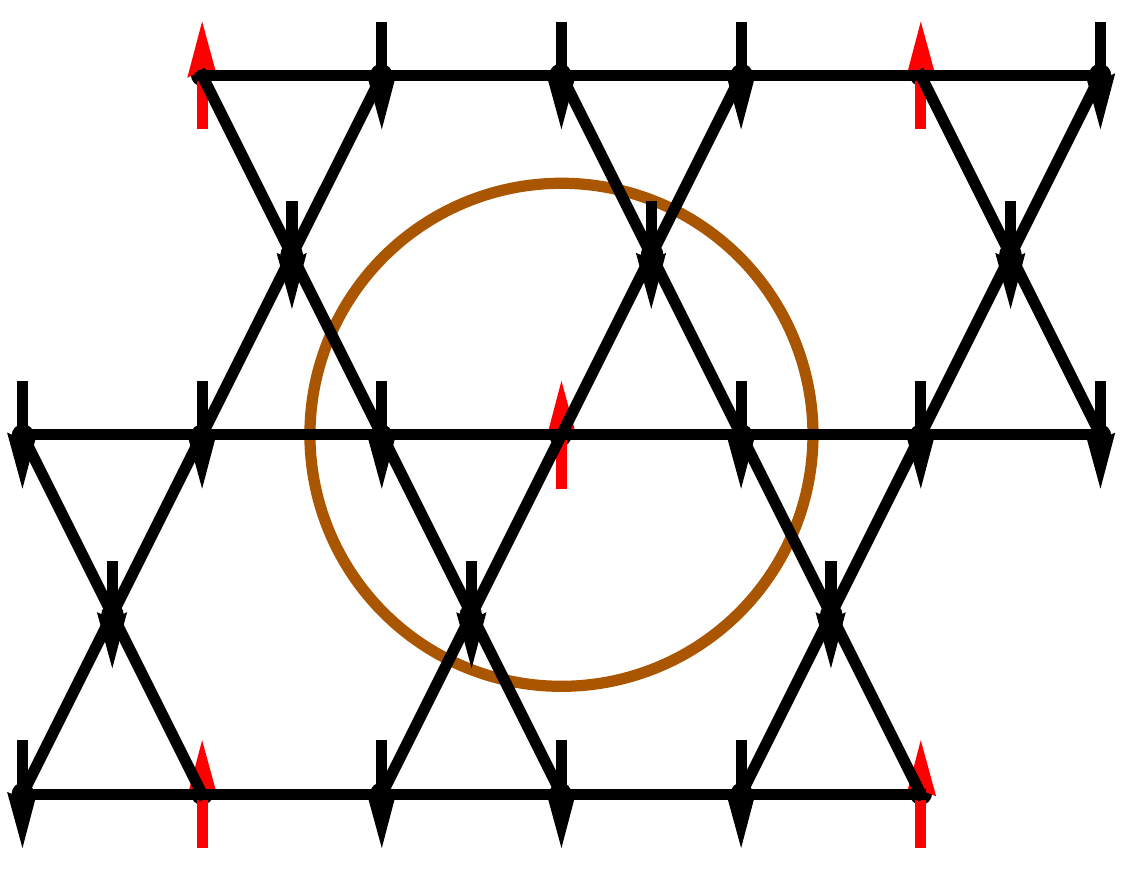}
       \subcaption{}
    \end{subfigure}
   \caption{(a) Blockade condition for the minima of $H_{\text{SPIM}}$ at its special point, $\delta k = 0$, $h \sim 0$. If the central spin is excited then its neighbours inside the red circle have to be unexcited. (b) Same for $H_{\text{150}}$. (c) Same for $H_{\text{kag}}$.}
    \label{fig:2Dconstrained}
\end{figure}

\tikzstyle{hackennode}=[draw,circle,fill=black,inner sep=0,minimum size=1.1pt]
\tikzstyle{hackenline}=[line width=0.17pt]

\section{Effective descriptions for other plaquette models}{\label{appendix:XORSATs}}

We can proceed as we did in Sec.~\ref{p6x}-\ref{LGTs} and give effective ``Rydberg blockade'' descriptions for other plaquette models. Like the qTPMz, 
all these effective models have special points with exponential-in-size degeneracy of classical minima (like the $k=-3J,h=0$ in the qTPMz), and whose quantum fluctuations in their vicinities give rise to blockaded dynamics and are possible candidates for spin liquids.

Consider the following two-dimensional systems with Hamiltonians
\begin{align}
   H_{\text{SPIM}} = &J \sum_{\{ i,j,k,l \} \in \tetragon} Z_i Z_j Z_k Z_l - (4+\delta k) \sum_i Z_{i} - h \sum_i X_i 
   \label{eq:2D1} \\
   H_{\text{150}} = &J \sum_{\{ i,j,k,l \} \in \tetrapleuro} Z_i Z_j Z_k Z_l - (4+\delta k) \sum_i Z_{i} - h \sum_i X_i
   \label{eq:2D2} \\
   H_{\text{kag}} = -&J\sum_{
         \{i, j, k\} \in \; \begin{tikzpicture}[baseline=-0.65ex,scale=0.16]
         \node[hackennode] (middle)   at ( 0,   0) {};
         \node[hackennode] (left)     at (-0.4,-0.6) {};
         \node[hackennode] (right)    at ( 0.4,-0.6) {};
         \node[hackennode] (topleft)  at (-0.4, 0.6) {};
         \node[hackennode] (topright) at ( 0.4, 0.6) {};
         \draw[hackenline,black]
            (left) -- (right);
         \draw[hackenline,black]
            (topleft) -- (topright);
         \draw[hackenline,black]
            (left) -- (topright);
         \draw[hackenline,black]
            (right) -- (topleft);
         \end{tikzpicture}}
      Z_i Z_j Z_k + (3+\delta k) \sum_i Z_{i} - h \sum_i X_i
   \label{eq:2D3}
\end{align}

Just like the qTPMz is built around the classical TPM (itself related to Rule 60 CA \cite{sfairopoulos2023boundary}), these are quantum models based on the classical square plaquette Ising model (SPIM) \cite{lipowski2000slow,jack2005caging,jack2005glassy}, the Rule 150 plaquette model \cite{sfairopoulos2023cellular}, and to the three-spin interaction model on the kagome lattice; see Fig.~\ref{fig:2Dconstrained}. In the expressions above the special point is at $\delta k=h=0$ in all cases. 

\begin{figure}
   \centering
   \includegraphics[width=0.6\columnwidth]{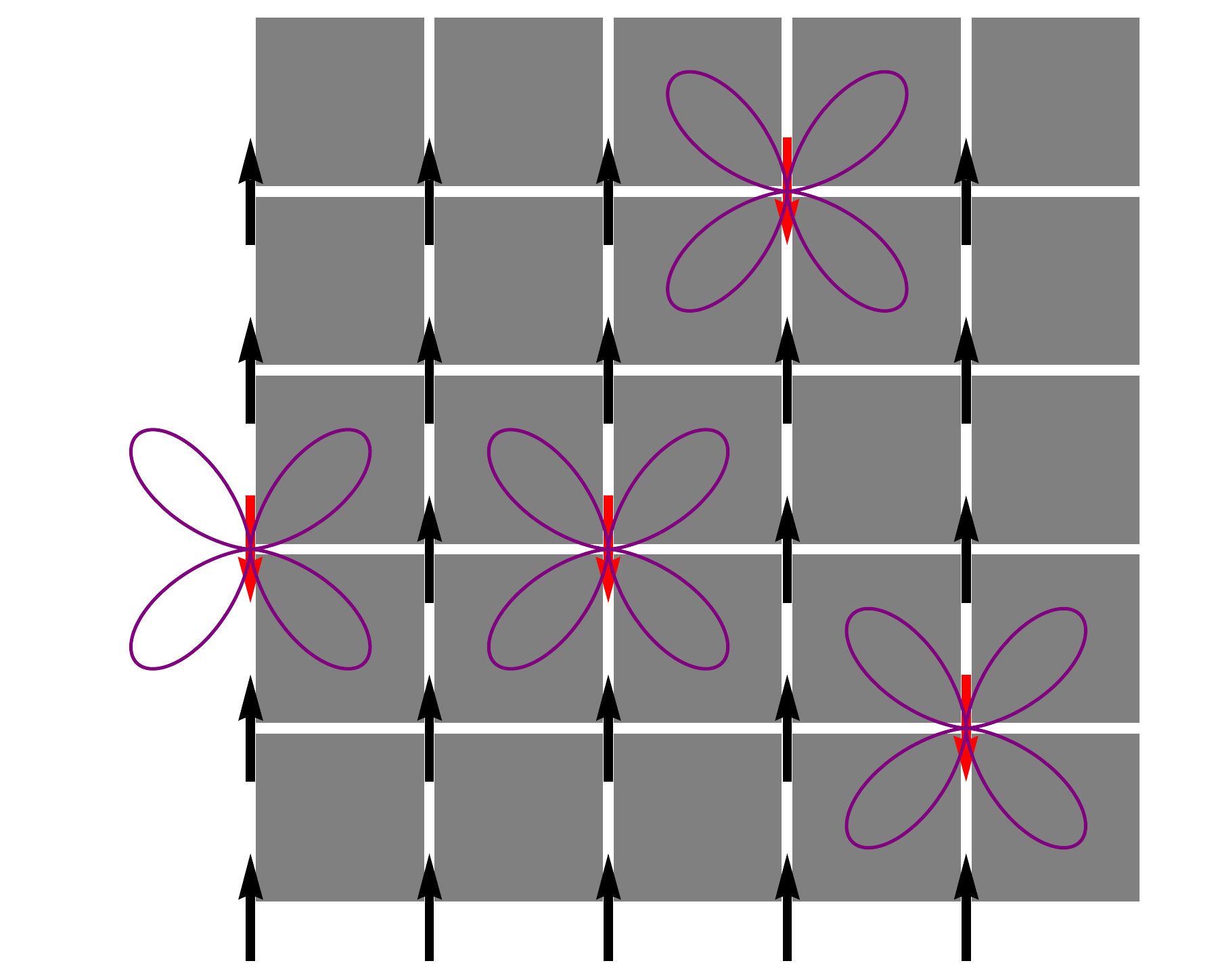}
   \caption{Tetramer description of the minima of $H_{\text{SPIM}}$ at its special point with PBC.}
    \label{fig:tetramers}
\end{figure}

Note that, although the SPIM and the Rule 150 both have 4-spin interactions, they give different shapes for their Rydberg blockade region with a different number of spins in them. Taking the SPIM as an example, the model in the conjectured spin liquid regime will involve a description in terms of {\em tetramers}, as shown in Fig.~\ref{fig:tetramers}: each site will be allowed to be excited only if none of its four neighboring dual lattice sites is occupied by a fermion. The same occurs for the other two models.

\begin{figure}
   \centering
   \includegraphics[width=0.6\columnwidth]{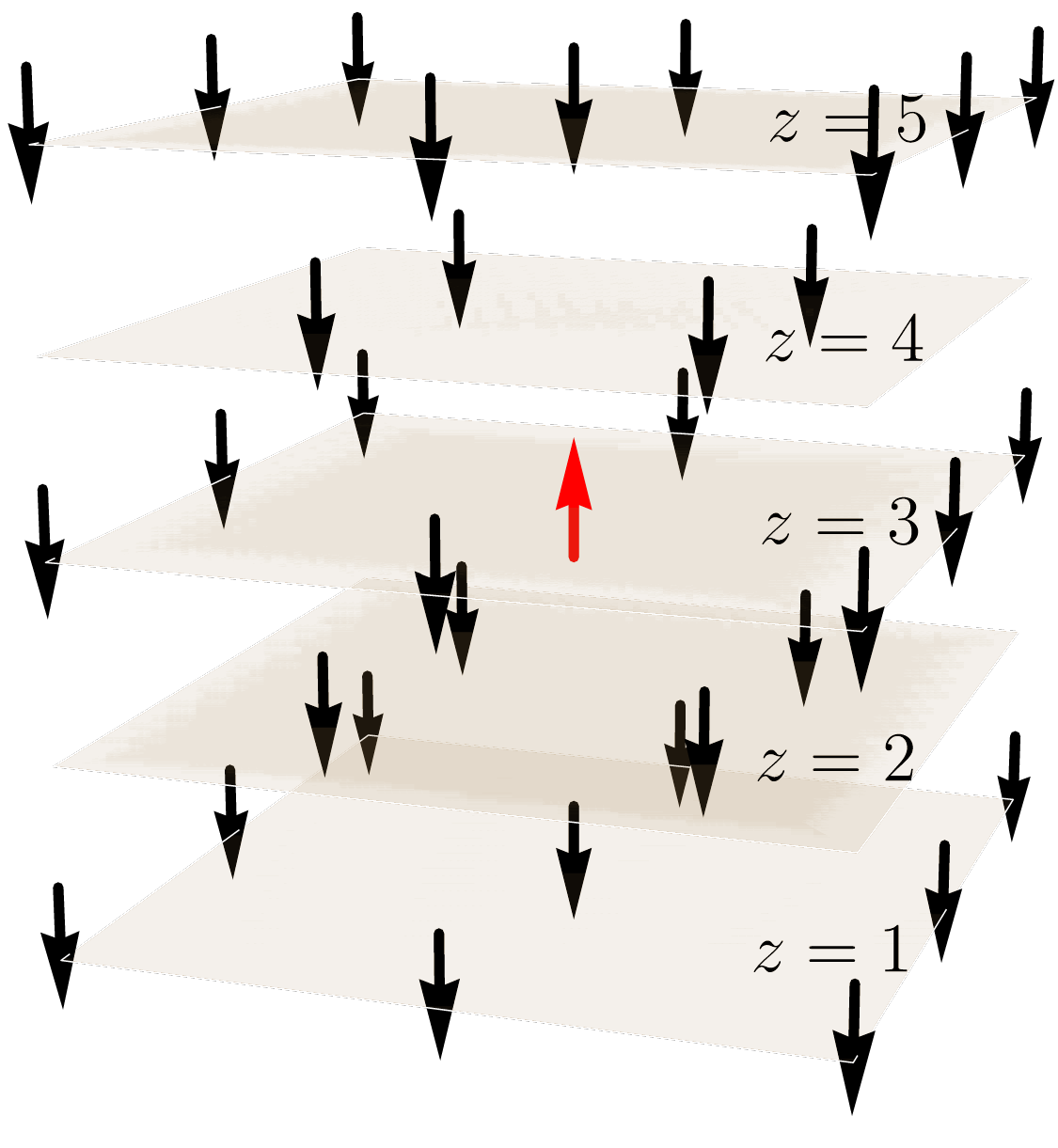}
   \caption{Same as Fig.~\ref{fig:2Dconstrained} but for \er{ESPyM}. The blockade involves spins from three different vertical levels ($z=2,3,4$ in the example), including 12 neighbors in total, 4 for each level.}
    \label{fig:SPyM constraint}
\end{figure}

A three-dimensional example of the same construct is that of the quantum square pyramid model (qSPyM), 
\tdplotsetmaincoords{70}{120}
\noindent
\begin{equation}
   H_{\text{qSPyM}} = - J\sum_{\{ i,j,k,l,m \} \in {\textstyle\mathstrut}  \raisebox{-0.3ex}{\begin{tikzpicture}[tdplot_main_coords,line cap=butt,line join=round,c/.style={circle,fill,inner sep=1pt},
      declare function={a=1.0/6;h=1.4/5;}]
      \path
      (0,0,0) coordinate (A)
      (a*2.3/2,-1/20,0) coordinate (B)
      (a,1.7*a/2,0) coordinate (C)
      (-0.0005,a*2.5/2,0) coordinate (D)
      (0,0,-h)  coordinate (S);
      \draw (S) -- (D) -- (C) -- (B) -- cycle (S) -- (C);
      \draw (B) -- (A) --(D);
      \draw[densely dotted] (A) -- (S);
  \end{tikzpicture}}}
  Z_i Z_j Z_k Z_l Z_m + (5 + \delta) \sum_i Z_{i} - h \sum_i X_i,
\label{ESPyM}
\end{equation}
related to the classical SPyM \cite{turner2015overlap,jack2016phase}. In this case, the blockade constraint implies that each spin is allowed to be excited if none of its five neihbouring dual sites are occupied by fermions, see Fig.~\ref{fig:SPyM constraint}. We can understand this from the following argument: each spin belongs to five downward-pointing pyramid terms, and in the center of mass of each of them we place a fermionic degree of freedom. Thus, flipping a single spin would affect the values of its five neighboring fermions.

From the above examples one can infer that all $p$-regular $p$-XORSAT models on the lattice (for example, for the TPM $p=3$), with the addition of a longitudinal field term per interaction plaquette, give rise to features similar to the $k=-pJ$ point for the TPMz. Note that not all of these constructions lead to exact constraint satisfaction problems (CSPs) in $\mathbb{F}_2$, but in general to more complex effective constraints in $\mathbb{F}_q$ with $q \geq 3$ with a constrained low-energy subspace as in the examples above.
The TPMz is an exception to this, as at $k=-3J$ its minima are given by a CSP with $q=2$, see \er{eq:km3}.
These models lead to equivalent formulations with respect to $p$-mers.

\end{document}